\documentclass[a4paper,11pt,notitlepage]{article}

\usepackage{geometry}
\usepackage{amsfonts}
\usepackage{amsmath,amsthm}
\usepackage{amssymb}
\usepackage{amsfonts}
\usepackage{amsmath,amsthm}
\usepackage{amssymb}
\usepackage{mathrsfs}
\usepackage{amsrefs}
\usepackage{color}
\usepackage{hyperref}
\definecolor{linkcolour}{rgb}{0,0.2,0.6}
\hypersetup{colorlinks,breaklinks,
            linkcolor=linkcolour,citecolor=linkcolour,
            filecolor=linkcolour, urlcolor=linkcolour}
\usepackage{tikz}
\usetikzlibrary{shapes.geometric}
\usetikzlibrary{decorations.markings}
\usepgflibrary{decorations.shapes}
\usepgflibrary{shapes.misc}

\newcommand{\ket}[1]{\left| #1 \right>}

\newcommand{\smfrac}[2]{\genfrac{}{}{}{1}{#1}{#2}}
\newcommand{\rectamup}[4]{\begin{tikzpicture}[]
    \useasboundingbox (-0.2,0) rectangle (0.8,0.7);
    \draw (0,0) -- (0.6,0); \draw (0,0.4) -- (0.6,0.4); 
    \draw (0,0) -- (0,0.4); \draw (0.6,0.4) -- (0.6,0); 
    \draw[fill=black] (0,0.4) circle (0.04); 
    \draw[fill=black] (0.6,0.4) circle (0.04); 
    \draw (0.3,-0.15) node{{\small $#1$}};
    \draw (0.3,0.55) node{{\small $#2$}};
    \draw (0.75,0.2) node{{\small $#3$}};
    \draw (-0.15,0.2) node{{\small $#4$}};
    \draw[fill=black] (0,0) circle (0.04); 
    \draw[fill=black] (0.6,0) circle (0.04); 
    \draw[->] (0.3,0.05) -- (0.3,0.35); 
  \end{tikzpicture}}
\newcommand{\rectamright}[4]{\begin{tikzpicture}[]
    \useasboundingbox (-0.2,0) rectangle (0.8,0.7);
    \draw (0,0) -- (0.6,0); \draw (0,0.4) -- (0.6,0.4); 
    \draw (0,0) -- (0,0.4); \draw (0.6,0.4) -- (0.6,0); 
    \draw[fill=black] (0,0.4) circle (0.04); 
    \draw[fill=black] (0.6,0.4) circle (0.04); 
    \draw (0.3,-0.15) node{{\small $#1$}};
    \draw (0.3,0.55) node{{\small $#2$}};
    \draw (0.75,0.2) node{{\small $#3$}};
    \draw (-0.15,0.2) node{{\small $#4$}};
    \draw[fill=black] (0,0) circle (0.04); 
    \draw[fill=black] (0.6,0) circle (0.04); 
    \draw[->] (0.05,0.2) -- (0.55,0.2); 
  \end{tikzpicture}}
\newcommand{\Zrect}[4]
{Z\left({\scriptstyle #4} \, {}^{#2}_{#1}\, {\scriptstyle #3}\right)}

\newcommand{\Ztree}{Z^{\mbox{\scriptsize{tree}}}}

\newcommand{\bsket}[3]{|{}_{#1 \phantom{a} #2}^{\phantom{a} #3}\rangle}
\newcommand{\bsbra}[3]{\langle{}^{#1 \phantom{a} #2}_{\phantom{a} #3}|}

\newcommand{\BffF}{|B_{ff}^+\rangle}

\usepackage{ctable}
\usepackage{array}

\begin{document}
\title{\textbf{Rectangular amplitudes, conformal blocks, and applications to loop models}}

\author{\\Roberto Bondesan$^{1,2}$, Jesper L.\ Jacobsen$^{1,3}$ 
  and Hubert Saleur$^{2,4}$\vspace{0.5cm} \\
  {\small ${}^1$LPTENS, \'Ecole Normale Sup\'erieure, 24 rue Lhomond,
  75231 Paris, France}\\ 
  {\small ${}^2$ Institute de Physique Th\'eorique, CEA Saclay, F-91191
  Gif-sur-Yvette, France}\\
  {\small ${}^3$Universit\'e Pierre et Marie Curie, 4 place Jussieu,
  75252 Paris, France} \\
  {\small ${}^4$ Physics Department, USC, Los Angeles, CA 90089-0484, USA} }

\pagestyle{headings} 

\date{}

 \maketitle
 \begin{abstract}
   \noindent 
   In this paper we continue the investigation of partition functions
   of critical systems on a rectangle initiated in [R.~Bondesan 
   \emph{et al}, Nucl.Phys.B862:553-575,2012].
   Here we develop a general formalism of rectangle boundary states
   using conformal field theory, adapted to describe geometries
   supporting different boundary conditions.
   We discuss the computation of rectangular amplitudes and their 
   modular properties, presenting explicit results for the case
   of free theories. In a second part of the paper we focus
   on applications to loop models, discussing in details
   lattice discretizations using both numerical and analytical 
   calculations. These results allow to interpret geometrically
   conformal blocks, and as an application we derive new probability
   formulas for self-avoiding walks.
 \end{abstract}



\addtolength{\baselineskip}{3pt}

\section{Introduction}
\label{sec:intro}

The study of conformal field theories (CFT) on a rectangle, while relying on a well known theoretical framework, offers interesting technical challenges, and had not been considered much in the literature until quite recently. In the last few years, the topic has however received  increased attention, both in the context of open string field theory\cite{Imamura2006,Imamura2007,Yin2007}, and in the context of condensed matter physics applications. The latter include study of quantum quenches for systems with open boundary conditions \cite{Calabrese2007,Dubail2011}, or the study of transport properties in disordered network models \cite{Cardy2000,Bondesan2012}. 

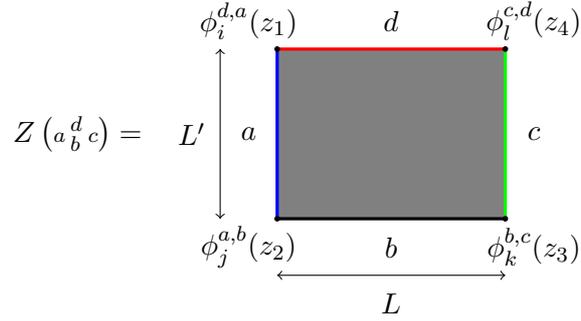
\begin{figure}
\begin{center}
\begin{tikzpicture}[scale=0.75]
  \node at (-5.5,1.5) {$\Zrect{b}{d}{c}{a}=$}; 
  \filldraw[gray] (-2,0) rectangle (2,3);
  \draw[black,very thick] (-2,0)--(2,0);
  \draw[red,very thick] (-2,3)--(2,3);
  \draw[blue,very thick] (-2,0)--(-2,3);
  \draw[green,very thick] (2,0)--(2,3);
  \draw[fill=black] (-2,0) circle (0.04);
  \draw[fill=black] (2,0) circle (0.04);
  \draw[fill=black] (-2,3) circle (0.04);
  \draw[fill=black] (2,3) circle (0.04);
  \node at (-2-0.5,1.5) {$a$};
  \node at (2+0.5,1.5) {$c$};
  \node at (0,-0.5) {$b$};
  \node at (0,3+0.5) {$d$};
  \node at (-2-0.5,-0.5) {$\phi_j^{a,b}(z  _2)$};
  \node at (-2-0.5,3+0.5) {$\phi_i^{d,a}(z_1)$};
  \node at (2+0.5,-0.5) {$\phi_k^{b,c}(z_3)$};
  \node at (2+0.5,3+0.5) {$\phi_l^{c,d}(z_4)$};
  \draw[<->] (-2,-1)--(2,-1);
  \node at (0,-1.5) {$L$};
  \draw[<->] (-3,0)--(-3,3);
  \node at (-3.5,1.5) {$L'$};
\end{tikzpicture}
\end{center}
\caption{The partition function for a CFT on a rectangle.
Different boundary conditions are imposed on each side, 
and boundary condition changing
operators are inserted at the corners $z_i$.}
\label{fig:rect_z}
\end{figure}

For a CFT on a rectangle, the topology is trivial, while there are sharp corners, and in general four different boundary conditions on the four edges, see figure \ref{fig:rect_z}. 

The most natural way to calculating properties---such as partition functions---in this geometry relies on a concept of ``fully open'' boundary state, which was introduced in \cite{Bondesan2011}. In the special case where all boundary conditions are  the same, this state was studied in details in \cite{Bondesan2011}, and related to remarkable algebraic properties of the Virasoro algebra. In the same reference, the correspondence between the CFT and lattice model descriptions of the rectangle geometry 
 was also considered in details. The present paper is a continuation of this work, where we now consider the case of different boundary conditions. 
 
The first two sections delineate the general formalism. In section \ref{sec:sol_gluing}  we introduce the most general fully open boundary states  $| B^b_{ac} \rangle$ and discuss their general expression in terms of  
 {\sl basis states} $\bsket{i}{j}{s}$, which are the natural analog  of the   Ishibashi states \cite{Ishibashi1988} in this context. In section  \ref{sec:ampl_cb}, we study amplitudes (partition functions), and compare the approaches using the  open boundary states with the approach using conformal mappings and conformal blocks. Various checks and related results are considered in section \ref{sec:free_theo} where we discuss  free theories. 
 
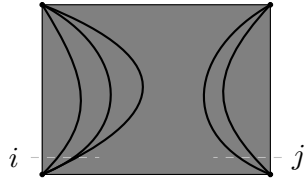
\begin{figure}
\begin{center}
\begin{tikzpicture}[scale=0.75]
  \filldraw[gray] (-2,0) rectangle (2,3);
  \draw [thick] (-2,0).. controls (-1,1) and (-1.2,1.7)
  ..(-2,3); 
  \draw [thick] (-2,0).. controls (-0.3,0.7) and (-0.5,2.2)
  ..(-2,3); 
  \draw [thick] (-2,0).. controls (0.2,1) and (0.5,2)
  ..(-2,3); 
  \draw [thick] (2,0).. controls (1,1) and (0.8,1.7)
  ..(2,3); 
  \draw [thick] (2,0).. controls (0.3,0.7) and (0.6,2.2)
  ..(2,3); 
  \draw[] (-2,0)--(2,0);
  \draw[] (-2,3)--(2,3);
  \draw[] (-2,0)--(-2,3);
  \draw[] (2,0)--(2,3);
  \draw[fill=black] (-2,0) circle (0.04);
  \draw[fill=black] (2,0) circle (0.04);
  \draw[fill=black] (-2,3) circle (0.04);
  \draw[fill=black] (2,3) circle (0.04);
  \draw[dashed,gray!50] (-2.2,0.3)--(-1,0.3);
  \node at (-2.5,0.3) {$i$};
  \draw[dashed,gray!50] (2.2,0.3)--(1,0.3);
  \node at (2.5,0.3) {$j$};
\end{tikzpicture}
\end{center}
\caption{A schematic illustration of the
  partition function for a loop model where we insert
$i$ lines in the bottom left corner and $j$ in the bottom right
one, and we impose that $s=i+j$ lines propagate vertically.}
\label{fig:part_fun_lines}
\end{figure}

 We then turn in section \ref{sec:lattice_states}  to our main  motivation for this work, which is the calculation of partition functions for loop models of statistical mechanics with lines inserted at the corners, for an example see figure \ref{fig:part_fun_lines}. These objects have potential relations with  transport properties of network models as well as potential probabilistic applications. They also provide a natural tool to investigate degeneracy issues when the corresponding conformal field theory becomes logarithmic \cite{Pearce2006,Read2007}. Although the conformal field theoretic description of loop models is not fully understood, we are able to find out the detailed connection between the general considerations of the first part of this paper and boundary conditions defined geometrically. This allows us to interpret conformal blocks in terms of loops, and calculate the desired partition functions. 
 We find in particular that coefficients in geometrical amplitudes  for percolation, dilute or dense polymers are then related with indecomposability parameters. 

A few final remarks are gathered in the conclusion, while some technical points are addressed in the appendices.

\section{Solutions of the gluing condition for semi-rectangular
geometries}
\label{sec:sol_gluing}

A boundary condition in conformal field theory has to preserve
conformal symmetry. This is encoded in the so-called gluing conditions
for the stress tensor of the CFT, gluing left- and right-moving modes.
In the familiar case of a theory defined on a disk, this implies that
\begin{equation}
  \label{eq:gluing_disk}
  \left(L_n-\bar{L}_{-n}\right)\left| B^p \right> = 0\, ,
\end{equation}
where $\left| B^p \right>$ is the boundary state describing the
boundary of the disk.  $\left| B^p \right>$ is an element of the space
of states $F$ of the bulk theory which we assume to decompose onto
Virasoro irreducibles $V_i$ with multiplicities $N_{ij}$,
$F=\bigoplus_{ij} N_{ij}V_i\otimes\bar{V}_j$.  The condition
\eqref{eq:gluing_disk} can be solved in each direct summand separately
and the unique solution in $V_i\otimes \bar{V}_i$ is given by
so-called Ishibashi states \cite{Ishibashi1988}:
\begin{equation}
  \label{eq:ishi}
  |i\rangle\rangle=
  \sum_N \left| i;N \right>\otimes  \overline{\left| i;N \right>}\, ,
\end{equation}
with 
$\left| i;N \right>$ ($\overline{\left| i;N
  \right>}$) an orthonormal base of $V_i$ ($\bar{V}_i$).

\begin{figure}[hpt]
\begin{center}
\begin{tikzpicture}[scale=1]
  \filldraw[gray] (0,0) rectangle (2,2.5);  
  \draw [very thick,red] (2,2.5) -- (2,0);
  \draw [very thick] (2,0) -- (0,0);
  \draw [very thick,blue] (0,0) -- (0,2.5) ;
  \draw[fill=black] (2,0) circle (0.05);
  \draw[fill=black] (0,0) circle (0.05);
  \node at (-0.25,-0.25) {$\phi_l$};
  \node at (2.25,-0.25) {$\phi_r$};
\end{tikzpicture}
\end{center}
\caption{The semi-rectangular geometry defining the boundary state.
}
\label{fig:bstate}
\end{figure}
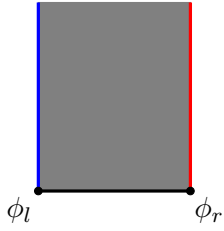

Using the transformation law of the stress tensor, also more
complicated boundaries can be described by gluing conditions.  In
\cite{Bondesan2011} the geometry of the bottom of the rectangle, see
figure \ref{fig:bstate}, was considered. In this case not only left
and right movers are glued together, but also positive and negative
modes within a chiral algebra. Further, conformal anomalies coming from
singularities of the stress tensor at the corners will show up in
the gluing condition.
Consider the possibility of having different boundary conditions whose
change is mediated by fields inserted in the left and right corners of
the bottom of the rectangle, of weights respectively $h_{l}$ and
$h_r$.  Then, denoting the boundary state $\left| B^o\right>$, the gluing
condition reads:
\begin{equation}
  \label{eq:gluing_rect}
  \left(L_n - L_{-n} -2n\left(\tilde{h}_l + (-)^n \tilde{h}_r\right) \right)
  \left| B^o\right> = 0 \, .
\end{equation}
where 
\begin{equation}
  \label{eq:h_tilde}
  \tilde{h} = 2h-\frac{c}{16}\, ,
\end{equation}
$c$ being the central charge.  

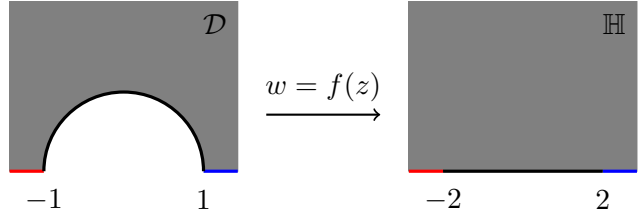
\begin{figure}
\begin{center}
\begin{tikzpicture}[scale=0.75]
	\begin{scope}
	\filldraw[gray] (-2,0) rectangle (2,3);
	\draw[very thick, white] (-1.4,0) -- (1.4,0);
	\draw[very thick,red] (-2,0) -- (-1.4,0); 
        \draw[very thick,blue] (1.4,0) -- (2,0);
	\draw (1.6,2.6) node{$\mathcal{D}$};
        \filldraw[white,smooth] (-1.4,0) arc (180:0:1.4);
	\draw[very thick] (-1.4,0) arc (180:0:1.4); 
        \node at (-1.4,-0.5) {$-1$};
        \node at (1.4,-0.5) {$1$};
	\end{scope}
        \node at (3.5,1.5) {$w=f(z)$};
	\draw[->,thick] (2.5,1) -- (4.5,1);
	\begin{scope}[xshift=7cm]
	\filldraw[gray] (-2,0) rectangle (2,3);
        \draw[very thick,red] (-2,0) -- (-1.4,0); 
        \draw[very thick,blue] (1.4,0) -- (2,0);
	\draw[very thick] (-1.4,0) -- (1.4,0);
	\draw (1.6,2.6) node{$\mathbb{H}$};
        \node at (-1.4,-0.5) {$-2$};
        \node at (1.4,-0.5) {$2$};
	\end{scope}
\end{tikzpicture}
\end{center}
\caption{Mapping of semicircular region $\cal{D}$ to the upper half plane
$\mathbb{H}$. $f$ is the Joukowsky map: $f(z)=z+z^{-1}$.}
\label{fig:map_D_UHP}
\end{figure}

The rectangle bottom in the $\zeta$-plane, with corners say in
$\zeta=-1,0$, can be mapped to the region
$\mathcal{D}=\{z\in\mathbb{C}: \Im(z)>0, |z|>1 \}$, left of figure
\ref{fig:map_D_UHP}, by $z=e^{-i\pi \zeta}$. This geometry is the
appropriate one to discuss boundary states in radial quantization,
and clearly the gluing condition (and thus the boundary state) is not
changed with respect to that for the bottom of the rectangle.
  In \cite{Bondesan2011} the operator implementing
algebraically the mapping from the upper half plane to the geometry
with corners $\mathcal{D}$, mapping $z=\pm 1$ to $w=\pm 2$, see figure
\ref{fig:map_D_UHP}, was found. It has the following product formula:
\begin{align}
  \label{eq:GD}
  \hat{G}_{\mathcal{D}} 
  = 
  \underset{N \rightarrow \infty}{{\rm lim}} e^{- \frac{1}{2^{N-1}} L_{-2^N}} 
  \dots e^{-\frac{1}{2} L_{-4}} e^{-L_{-2}} \, .
\end{align}

It follows that when $h_{l}=h_r=0$, homogeneous boundary conditions, the
boundary state is:
\begin{align}
  \label{eq:Bstate_hom}
  \left| B^o \right> 
  &= 
  \hat{G}_{\mathcal{D}} \left| 0\right> \\
  &= 
  | 0\rangle
  -
  L_{-2} | 0\rangle
  -
  \frac{1}{2} L_{-4} | 0\rangle 
  +
  \frac{1}{2} L_{-2}^2 | 0\rangle + \dots \, .
\end{align}
We will comment about the normalization
of the state $|B^o\rangle$ when we compute the amplitudes
involving boundary states below. In the following we will continue
to use the normalization fixed in equation \eqref{eq:Bstate_hom}.

In this paper we focus on the case of different boundary conditions on
the sides of the semi-rectangle.  To make
things concrete, call the condition on the left side $a$, that on the
bottom $b$ and that on the right $c$.
Boundary condition changing (BCC) operators $\phi_i^{a|b}$,
$\phi_j^{b|c}$---which can be identified with the lowest eigenstate of the
Hamiltonian on a strip with boundary conditions $a$ and $b$ on the
sides ---will then sit at the left and right corners.
We denote the boundary state we want to determine $| B^b_{ac} \rangle$,
which is thus associated graphically to:
\begin{equation}
  \label{eq:ish_pic}
  | B^b_{ac} \rangle=
  \begin{tikzpicture}[]
    \useasboundingbox (-0.3,0) rectangle (1,0.7);
    \draw (0,0) -- (0.8,0); 
    \draw (0,0) -- (0,0.6);
    \draw (0.8,0.6) -- (0.8,0); 
    \draw[fill=black] (0,0) circle (0.04); 
    \draw[fill=black] (0.8,0) circle (0.04); 
    \draw (-0.25,0.3) node{$a$};
    \draw (0.4,0.-0.25) node{$b$};
    \draw (1.05,0.3) node{$c$};
  \end{tikzpicture}
\end{equation}
If $X=\Phi(z_1,\bar{z}_1)\cdots \phi(x)\cdots$ is a chain of arbitrary bulk
and boundary operators, $\left| B^b_{ac} \right>$
is defined by
\begin{align}
  \label{eq:B_vac_map}
  \langle X  \rangle_{\mathcal{D}}=  
  \langle 0|X | B^b_{ac} \rangle\, ,
\end{align}
where $\langle X  \rangle_{\mathcal{D}}$ is the correlator in the geometry
$\mathcal{D}$.
The operator $\hat{G}_{\mathcal{D}}$ introduced above
by definition implements
at the operator level the conformal map from
$\mathcal{D}$ to $\mathbb{H}$, so that we have:
\begin{align}
  \langle X \rangle_{\mathcal{D}}
  =
  \langle 0 | \tilde{X} \phi_i^{a|b}(-2)\phi_j^{b|c}(2) |0\rangle\, ,
\end{align}
where we defined $\tilde{X}=\hat{G}^{-1}_{\mathcal{D}}  X \hat{G}_{\mathcal{D}} $
(if $\hat{G}_{\mathcal{D}} =e^{\sum_n \epsilon_n L_n}$, then $\hat{G}^{-1}_{\mathcal{D}} 
=e^{-\sum_n \epsilon_n L_n}$), and the right term is
a correlation function on the upper half plane. Since the
out-vacuum $\left< 0\right|$ is invariant under this mapping, one has
\begin{equation}
  \label{eq:Blr}
  | B^b_{ac} \rangle=\hat{G}_{\mathcal{D}} \phi_i^{a|b}(-2)\phi_j^{b|c}(2) 
  |0\rangle\, .
\end{equation}

We now recall the boundary OPE of two boundary fields 
(see e.g.~\cite{Lewellen1992}, $x>y$):
\begin{equation}
  \label{eq:bdry_OPE}
  \phi_i^{a,b}(x)\phi_j^{b,c}(y) 
  =  \sum_s (x-y)^{h_s-h_i-h_j}
  C_{ijs}^{abc} \phi^{a,c}_s \left(\smfrac{x+y}{2}\right)+\dots
  \, .
\end{equation}
The boundary
OPE coefficients $C_{ijs}^{abc}$ are part of the known data of the CFT
\cite{Runkel1999,Felder2000}:
\begin{equation}
  \label{eq:C_OPE_F}
  C_{ijs}^{abc}=F_{bs}{\scriptsize \left[ \begin{array}{cc}
           a & c \\
           i & j \end{array} \right]}\, ,
\end{equation}
where the $F$-matrices relate conformal blocks in different bases
\cite{Moore1990}. We will see explicit examples below.
Note that from taking different OPEs of the three point function
$\langle \phi_i^{ab}\phi_j^{bc}\phi_k^{ca}\rangle$ one has the cyclic relation
\begin{equation}
  C_{ijk}^{abc}C_{kk0}^{aca}=C_{jki}^{bca}C_{ii0}^{aba}\, .
\end{equation}
Further, the normalization of the identity $\phi_0$ ($0$ is the label
for the identity) is fixed by
demanding $\phi_0^{aa}\phi_{i}^{ab}= \phi_{i}^{ab} $, which implies
\begin{equation}
  \label{eq:c0ii}
  C_{0ij}^{aab}=\delta_{ij}\, ,
\end{equation}
for every $a,i$. 
More properties can be found in \cite{Lewellen1992,Runkel1999}.

Eq.~\eqref{eq:Blr} can then be written as:
\begin{align}
  \label{eq:sol_ish}
  | B^b_{ac} \rangle=
  \sum_s C_{ijs}^{abc}\sqrt{C_{ss0}^{aca}}
  \bsket{i}{j}{s}
  \, ,
\end{align}
where we defined the {\sl basis states} $\bsket{i}{j}{s}$ by
\begin{align}
  \label{eq:ish_state}
  \bsket{i}{j}{s} &:= \hat{G}_{{\mathcal D}} \; 4^{h_s-h_i-h_j}  
  \sum_{n\ge 0} 
  \sum_{\gamma \vdash n} 4^n
  \beta_{i,j}^{s,\gamma}
  L_{-\gamma} |\phi_s\rangle \\
  \label{eq:ish_state2}
  &=
   4^{h_s-h_i-h_j} \Big[
  |\phi_s \rangle 
  + 
  \sum_{n\ge 1} \sum_{\gamma \vdash n} 
  \delta_{i,j}^{s,\gamma}
  L_{-\gamma}
  |\phi_s \rangle \Big]\, ,
\end{align}
where $\gamma \vdash n$ is a partition $(\gamma_1,\dots,\gamma_L)$ of
$n$, and we used the shorthand $L_{-\gamma}:= L_{-\gamma_1}\cdots
L_{-\gamma_L}$.  The coefficients $\beta_{i,j}^{s,\gamma}$ appear in
the OPE of (bulk) chiral fields $\phi_{i},\phi_{j}$ in the term
corresponding to the descendant of $s$, $L_{-\gamma}\phi_s$
(We use the notations of \cite{DiFrancesco1997}, section 6.6.3 of this
reference, although we simplify the notation by writing $\gamma$
instead of $\{\gamma\}$). The term $\sqrt{C_{ss0}^{aca}}$ in
eq.~\eqref{eq:sol_ish} is needed in order to define the basis state as
an element of the Verma module of the primary $\phi_s$ without any
information about the boundary conditions.
Alternatively one could remove the factor $\sqrt{C_{ss0}^{aca}}$ and
define the basis state as an element of the Verma module of
$\phi^{a,c}_s$, where $\langle \phi^{a,c}_s|
\phi^{c,a}_s\rangle=C_{ss0}^{aca}$.

The basis states $\bsket{i}{j}{s}$ span the space of boundary states
solving eq.~\eqref{eq:gluing_rect} and are the analogues for the
rectangular geometry of Ishibashi states \eqref{eq:ishi}.  For
determining the coefficients $\delta$'s in eq.~\eqref{eq:ish_state2}
one could work out the algebra in \eqref{eq:ish_state} or directly
solve the constraint \eqref{eq:gluing_rect} for different values of
$n$. It is well-known however that the combinatorics of the Virasoro
algebra is complicated and does not allow to get a closed formula for
the coefficients in the above expansions, be it
$\beta_{i,j}^{s,\gamma}$ or $\delta_{i,j}^{s,\gamma}$. An exception is
the case of free theories, when the Virasoro generators admit an
expression in terms of bosonic or fermionic oscillators, see the
discussion in section \ref{sec:free_theo}.  In general, there is therefore  
the following 
difference between Ishibashi states for the cylinder geometry and the
basis states for the rectangle: rectangle basis states depend on the
boundary operators inserted at the corners, so on their fusion, which
does not give access to an explicit expression. Instead Ishibashi
states are known explicitly from eq.~\eqref{eq:ishi}.

The state $| B^b_{ac}\rangle$ is directly given by a single
basis state if the fusion $\phi_i^{a,b}\otimes\phi_j^{b,c}$ produces
only one channel.
This is of course the case of homogeneous boundary conditions,
when no BCC operators are present,
\begin{equation}
  \label{eq:Bhom}
  |B^a_{aa}\rangle=(C_{000}^{aaa})^{3/2} \bsket{0}{0}{0}
  = \bsket{0}{0}{0}
  = \hat{G}_{{\mathcal D}}|0\rangle \, ,
\end{equation}
where 
$|\phi_0\rangle\equiv|0\rangle$, and we used \eqref{eq:c0ii}.
This is also the case when there is only a single
BCC operator. Indeed, if $a=b$, calling the BCC operator
$\phi_s^{a,c}$, one has
\begin{align}
  \label{eq:Baac}
  |B^a_{ac}\rangle 
  &=  C_{0ss}^{aac}\sqrt{C_{ss0}^{aca}}
  \bsket{0}{s}{s}\\
  &= \sqrt{C_{ss0}^{aca}}\hat{G}_{{\mathcal D}}\phi_s(2)|0\rangle
  = \sqrt{C_{ss0}^{aca}} \hat{G}_{{\mathcal D}}\exp(2L_{-1})|\phi_s\rangle \, ,
\end{align}
since $L_{-1}$ generates translations and $|\phi_s\rangle$ inserts the
operator at the origin.  $C_{ss0}^{aca}$ is not fixed
by \eqref{eq:c0ii} and could be set to one, see \cite{Runkel1999}.

The discussion presented so far solves
implicitly the problem of finding the boundary states for the
rectangular geometry.  In the next section we will compute amplitudes
involving these states and give more explicit results. Before moving
on, let us define the conjugate states of eq.~\eqref{eq:sol_ish}
\begin{align}
  \label{eq:sol_ish_dual}
  \langle B^b_{ac} |=
  \sum_s C_{ijs}^{abc}\sqrt{C_{ss0}^{aca}}
  \bsbra{i}{j}{s}
  \, ,
\end{align}
which are associated graphically to 
\begin{equation}
  \langle B^b_{ac} |=
  \begin{tikzpicture}[]
    \useasboundingbox (-0.3,0) rectangle (1,0.7);
    \draw (0,0.6) -- (0.8,0.6); 
    \draw (0,0) -- (0,0.6);
    \draw (0.8,0.6) -- (0.8,0); 
    \draw[fill=black] (0,0.6) circle (0.04); 
    \draw[fill=black] (0.8,0.6) circle (0.04); 
    \draw (-0.25,0.3) node{$a$};
    \draw (0.4,0.85) node{$b$};
    \draw (1.05,0.3) node{$c$};
  \end{tikzpicture}\, .
\end{equation}
$\bsbra{i}{j}{s}$ is the conjugate of $\bsket{i}{j}{s}$, and we 
introduce also the following graphical notation for the basis
states:
\begin{equation}
  \bsket{i}{j}{s}=
  \begin{tikzpicture}[]
    \useasboundingbox (-0.3,0) rectangle (1,0.7);
    \draw (0,0) -- (0.8,0); 
    \draw (0,0) -- (0,0.6);
    \draw (0.8,0.6) -- (0.8,0); 
    \draw[fill=black] (0,0) circle (0.04); 
    \draw[fill=black] (0.8,0) circle (0.04); 
    \draw (-0.15,-0.25) node{$\phi_i$};
    \draw (0.95,-0.25) node{$\phi_{j}$};
    \draw (0.4,0.4) node{$s$};
  \end{tikzpicture}
\end{equation}

\section{Amplitudes and conformal blocks}
\label{sec:ampl_cb}

We consider a rectangle of length $L'$ and width $L$ in the $z$-plane,
with boundary conditions $a,b,c,d$, so that BCC operators
$\phi^{a,d}_i(z_1),\phi^{a,b}_j(z_2),\phi^{b,c}_k(z_3),\phi^{d,c}_l(z_4)$
sit at the corners, see figure \ref{fig:rect_z}.
The CFT partition function is by definition:
\begin{align}
  \label{eq:Z_rect_bcc}
  \Zrect{b}{d}{c}{a}=\rectamup{b}{d}{c}{a} = 
  Z_0(\tau) 
  \langle \phi^{a,d}_i(z_1),\phi^{a,b}_j(z_2),\phi^{b,c}_k(z_3),\phi^{d,c}_l(z_4)
  \rangle\, ,
\end{align}
where $Z_0$ is the partition function for homogeneous boundary conditions and
the correlator is evaluated on the rectangle. The arrow in the picture
is the direction of imaginary time in a transfer matrix picture, which
will be developed below, and
$\tau=iL'/L$.

We recall first that $Z_0$
has been computed in \cite{Kleban1992}:
\begin{equation}
  \label{eq:Z0}
  Z_0(\tau) = \rectamup{a}{a}{a}{a} = L^{c/4} \eta^{-c/2}(\tau) \, ,
\end{equation}
where $\eta$ is the Dedekind eta function. Definitions and properties
of the $\eta$ function are collected in appendix \ref{sec:usef_form}.
In \cite{Bondesan2011},
we showed that the following relation between the boundary state
\eqref{eq:Bstate_hom} and this partition function exists:
\begin{equation}
  \label{eq:B0eta}
  \langle B^o|
  \hat{q}^{L_0-c/24}  |B^o \rangle = L^{-c/4}Z_0 =\eta(\tau)^{-c/2}\, ,
\end{equation}
$\hat{q}$ being the relevant combination of $L'/L$
appearing in the strip transfer matrix:
\begin{equation}
  \label{eq:q_tau}
  \hat{q}=\sqrt{q}=\exp(\pi i \tau) \, .
\end{equation}
From relation \eqref{eq:B0eta} we note that the state
$|B^o\rangle$ is not normalizable:
\begin{equation}
  \langle B^o|B^o\rangle = \lim_{q\to 1}\eta(\tau)^{-c/2}=
  \begin{cases} 
    \infty & \text{if $c>0$,}\\
    0 &\text{if $c< 0$,}\\
    1 &\text{if $c= 0$.}
  \end{cases}
\end{equation}
A similar situation happens for the Ishibashi states in the cylinder
geometry. However in the case of cylinder states, one can fix 
a possible multiplicative
constant in the definition, which is not fixed by the gluing condition
on the disk \eqref{eq:gluing_disk}, by requiring the
equality of the cylinder amplitude to the annulus partition function,
see \cite{Cardy1989}.
We do not know how to fix a possible multiplicative constant $\alpha$
in the definition $|B^o\rangle = \alpha \hat{G}_{\mathcal{D}} \left| 0\right>$.
This issue is related to the very definition of a partition function on 
a rectangle, containing the factor  $L^{c/4}$ 
(or $L^w$ in the general case, see below)
in \eqref{eq:Z0}, which depends
on the unit of measure of length we adopt.


We now go back to the computation of \eqref{eq:Z_rect_bcc}.
We choose the following fusion channel of BCC operators
(here depicted on a segment):
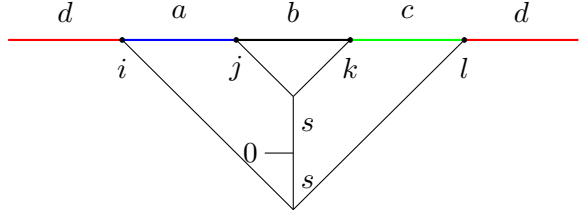
\begin{center}
\begin{tikzpicture}[scale=0.75]
  \draw[thick, blue] (-3,0)--(-1,0);
  \draw[thick, black] (-1,0)--(1,0);
  \draw[thick, green] (1,0)--(3,0);
  \draw[thick, red] (-5,0)--(-3,0);
  \draw[thick, red] (3,0)--(5,0);
  \draw[fill=black] (-3,0) circle (0.04); 
  \draw[fill=black] (-1,0) circle (0.04); 
  \draw[fill=black] (1,0) circle (0.04); 
  \draw[fill=black] (3,0) circle (0.04); 
  \node at (-3,-0.5) {$i$};
  \node at (-1,-0.5) {$j$};
  \node at (1,-0.5) {$k$};
  \node at (3,-0.5) {$l$};
  \node at (-4,0.5) {$d$};
  \node at (-2,0.5) {$a$};
  \node at (0,0.5) {$b$};
  \node at (2,0.5) {$c$};
  \node at (4,0.5) {$d$};
  \draw (-1,0)--(0,-1);
  \draw (1,0)--(0,-1);
  \draw (-3,0)--(0,-3);
  \draw (3,0)--(0,-3);
  \draw (0,-1)--(0,-3);
  \draw (0,-2)--(-0.5,-2);
  \node at (-0.75,-2) {$0$};
  \node at (0.25,-2.5) {$s$};
  \node at (0.25,-1.5) {$s$};
\end{tikzpicture}
\end{center}
This corresponds to fuse fields along the bottom and top segments
of the rectangle.
The partition function has the 
following expression in a transfer matrix formulation in
terms of boundary states:
\begin{align}
  \label{eq:Z_ampli}
  \Zrect{b}{d}{c}{a}=\rectamup{b}{d}{c}{a} &= 
  L^{w(i,j,k,l)}
  \langle B_{ac}^d|\hat{q}^{L_0-c/24}| B_{ac}^b\rangle\\
  \label{eq:Z_ampli2}
  &=
  L^{w(i,j,k,l)}
  \sum_s C_{ils}^{adc}C_{jks}^{abc}C_{ss0}^{aca}
  \bsbra{i}{l}{s}
  \hat{q}^{L_0-c/24}
  \bsket{j}{k}{s}\, .
\end{align}
The modular weight $w(i,j,k,l)$ appearing in the term $ L^{w(i,j,k,l)}$ is
given by the anomalies at the corners \cite{Cardy1988}:
\begin{equation}
  \label{eq:mod_weight}
  w(i,j,k,l) = \frac{c}{4} - 2 (h_i+h_j+h_l+h_k) \, .
\end{equation}

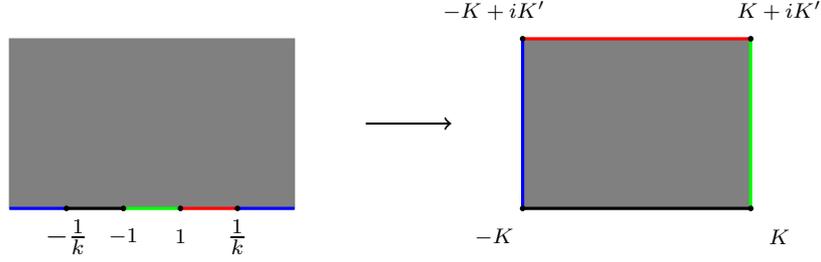
\begin{figure}[hpt]
\centering
\begin{tikzpicture}[scale=0.75]
    \filldraw[gray] (-2,0) rectangle (3,3);
    \draw[very thick,blue] (-2,0) -- (-1,0); 
    \draw[very thick,black] (-1,0) -- (0,0);
    \draw[very thick,green] (0,0) -- (1,0);
    \draw[very thick,red] (1,0) -- (2,0);
    \draw[very thick,blue] (2,0) -- (3,0);
  \draw[fill=black] (-1,0) circle (0.04);
  \draw[fill=black] (0,0) circle (0.04);
  \draw[fill=black] (1,0) circle (0.04);
  \draw[fill=black] (2,0) circle (0.04);
    \node at (-1,-0.5) {$-\smfrac{1}{k}$};
    \node at (0,-0.5) {{\scriptsize $-1$}};
    \node at (1,-0.5) {{\scriptsize $1$}};
    \node at (2,-0.5) {$\smfrac{1}{k}$};
  \draw[->,thick] (4.25,1.5) -- (5.75,1.5);
  \begin{scope}[xshift=9cm]
  \filldraw[gray] (-2,0) rectangle (2,3);
  \draw[black,very thick] (-2,0)--(2,0);
  \draw[red,very thick] (-2,3)--(2,3);
  \draw[blue,very thick] (-2,0)--(-2,3);
  \draw[green,very thick] (2,0)--(2,3);
  \draw[fill=black] (-2,0) circle (0.04);
  \draw[fill=black] (2,0) circle (0.04);
  \draw[fill=black] (-2,3) circle (0.04);
  \draw[fill=black] (2,3) circle (0.04);
  \node at (-2-0.5,-0.5) {{\scriptsize $-K$}};
  \node at (2+0.5,-0.5) {{\scriptsize $K$}};
  \node at (2+0.5,3+0.5) {{\scriptsize $K+iK'$}};
  \node at (-2-0.5,3+0.5) {{\scriptsize $-K+iK'$}};
  \end{scope}
\end{tikzpicture}
    \caption{The Schwarz-Cristoffel transformation of eq.~\eqref{eq:SCT}
    sending the the upper half plane to the rectangle.}
    \label{fig:SCT}
\end{figure}

The building blocks of the partition function \eqref{eq:Z_ampli2} are 
amplitudes involving basis states:
\begin{align}
  \label{eq:ampli_ish}
  \mathcal{A}_{s}^{i,j,k,l}(\tau)
  = 
  \begin{tikzpicture}[]
   \useasboundingbox (-0.3,0) rectangle (1.0,1);
    \draw (0,0) -- (0.6,0); 
    \draw (0,0.4) -- (0.6,0.4); 
    \draw (0,0) -- (0,0.4);
    \draw (0.6,0.4) -- (0.6,0); 
    \draw[fill=black] (0,0.4) circle (0.04); 
    \draw[fill=black] (0.6,0.4) circle (0.04); 
    \draw (0.3,0.2) node{$s$};
    \draw[fill=black] (0,0) circle (0.04); 
    \draw[fill=black] (0.6,0) circle (0.04); 
    \draw (-0.15,0.65) node{$\phi_i$};
    \draw (0.75,0.65) node{$\phi_{l}$};
    \draw (-0.15,-0.25) node{$\phi_j$};
    \draw (0.75,-0.25) node{$\phi_{k}$};
    \draw[->] (0.8,0) -- (0.8,0.4); 
  \end{tikzpicture}
  :&= 
  \bsbra{i}{l}{s}
  \hat{q}^{L_0-c/24}
  \bsket{j}{k}{s}\\
  \label{eq:ampli_ish_ser}
  &=
  \hat{q}^{h_s-c/24}\sum_{n\ge 0}c_n \hat{q}^n
  \, .
\end{align}
In practice using the expression for boundary states
\eqref{eq:ish_state} one can only obtain
the first few coefficients $c_n$ of the power series above.  However
closed expressions can be computed using
CFT correlators of the BCC operators, since by definition one has the
following relation:
\begin{equation}
  \label{eq:AZ}
  L^{w(i,j,k,l)}
  \mathcal{A}_{s}^{i,j,k,l}(\tau)
  =
  Z_0(\tau) 
  \langle \phi_i(z_1)\phi_j(z_2)\phi_k(z_3)\phi_l(z_4)
  \rangle_s\, ,
\end{equation}
where only the channel $s$ in the correlator is considered.
For computing a correlator of the fields on the rectangle, we map it
to the upper half plane $w$
\begin{equation}
  \label{eq:phizphiw}
  \left< \phi_1(z_1) \cdots \phi_n(z_n)\right>
  =
  (z'(w_1))^{-h_1}\cdots (z'(w_n))^{-h_n}
  \left< \phi_1(w_1) \cdots \phi_n(w_n)\right>\, .
\end{equation}
This mapping is given by a Schwarz-Cristoffel transformation:
\begin{equation}
  \label{eq:SCT}
  z = \int_0^w {\rm d}t \frac{1}{\sqrt{(1-t^2)(1-k^2t^2)}}\, .
\end{equation}
It maps the points on the real line $(-1/k, -1, 1, 1/k)$ to the
corners of a rectangle with $L=2K$ and $L'=K'$, $(-K + iK', -K, K, K +
iK')$, where $K$ is the complete elliptic integral of the first kind
of modulus $k$ and $K'$ the complementary one, see figure \ref{fig:SCT}.

The Jacobian $J$ of this transformation is singular at the corners,
but when the boundary operators sit at the corners too,
eq.~\eqref{eq:phizphiw} continues to hold since as the argument of the
field approaches the corner in $z_c$, the field itself goes to zero as
$(z-z_c)^h \phi(z_c)$, where $h$ is the dimension of the field. Then
we look at the Jacobian as
the coefficient of the first term in the expansion around the position
of the corners.  The result is
\begin{equation}
  \label{eq:J_SCT}
  J= k^{-h_i-h_l}(1-k^2)^{h_i+h_j+h_k+h_l}\, .  
\end{equation}
Thanks to global conformal invariance, a chiral correlator 
can be written as \cite{DiFrancesco1997}
\begin{equation}
  \label{eq:4point_UHP}
  \langle \phi_1(w_1)\phi_2(w_2)\phi_3(w_3)\phi_4(w_4) \rangle
  =
  \prod_{i<j}w_{ij}^{\mu_{ij}} G(\zeta)\, ,\quad
  \mu_{ij}=\frac{1}{3}\left(\sum_{k=1}^4h_k\right) - h_i - h_j\, ,
\end{equation}
with $\zeta$ being the anharmonic ratio ($w_{ij}=w_i-w_j$)
\begin{equation}
  \label{eq:cross_ratio}
  \zeta=\frac{w_{12}w_{34}}{w_{13}w_{24}}\, .
\end{equation}
We now let $(w_1,w_2,w_3,w_4)=(-1/k, -1, 1, 1/k)$ and we find
\begin{align}
  \label{eq:pref}
  \prod_{i<j}w_{ij}^{\mu_{ij}} 
  =
  \left( k^{4(h_1+h_4)+h_2+h_3}\left(2(1-k^2)\right)^{-(h_1+h_2+h_3+h_4)} \right)^{1/3}\, .
\end{align}
Expressing everything in terms of $\tau$, using
eq.~\eqref{eq:K_theta}, we have:
\begin{equation}
  \label{eq:4point_rect}
  \langle \phi_{i}(z_1)\phi_{j}(z_2)\phi_{k}(z_3)\phi_{l}(z_4)\rangle
  = 
  \left(
  2^{\frac{1}{3}}\pi^2
  \eta^{4}
  (2K)^{-2}
  \right)^{h_i+h_j+h_k+h_l}
  G(\zeta)\, .
\end{equation}
We recognize in the above formula the term length $L=2K$ raised to the
correct modular weight \eqref{eq:mod_weight} coming from field
insertions. $G(\zeta)$ in general is a linear combination of conformal
blocks $\mathcal{F}_{il;jk}^{s}(\zeta)$, corresponding to the fusion channel $s$,
and in the end eq.~\eqref{eq:ampli_ish} becomes:
\begin{equation}
  \label{eq:ish_cb_prop}
  \mathcal{A}_{s}^{i,j,k,l}(\tau)
  \propto
  \eta(\tau)^{-2w(i,j,k,l)}
  \mathcal{F}_{il;jk}^{s}(1-\zeta(\tau)) \, ,
\end{equation}
where the anharmonic ratio as a function of $\tau$ is 
\begin{equation}
  \label{eq:anratio}
  \zeta = \left(\frac{k-1}{k+1}\right)^2 = 
  \left(\frac{\theta_4(\tau)}{\theta_3(\tau)}\right)^4 \, . 
\end{equation}
Let us
compute the constant of proportionality by demanding the normalization
of the amplitude to be $c_0=4^{2h_s-(h_i+h_j+h_k+h_l)}$ in
eq.~\eqref{eq:ampli_ish_ser}, according to eq.~\eqref{eq:ish_state2}.
When $q\to 0$, $1-\zeta(\tau)\to 0$, and using that our conventions
are such that
\begin{equation}
  \label{eq:norm_cb}
  \mathcal{F}_{il;jk}^{s}(z)\sim z^{h_s-\frac{1}{3}(h_i+h_j+h_k+h_l)}\, ,
\end{equation}
we have the following small-$q$ expansion of the right hand side of
\eqref{eq:ish_cb_prop}:
\begin{equation}
  \label{eq:small_q_am_right}
  \eta(\tau)^{-2w(i,j,k,l)}
  \mathcal{F}_{il;jk}^{s}(1-\zeta(\tau)) 
  =
  4^{2h_s-\frac{2}{3}(h_i+h_j+h_k+h_l)}
  q^{-c/48+h_s/2}(1+O(\hat{q}))
  \, .
\end{equation}
Then the exact relation is:
\begin{equation}
  \label{eq:ish_cb}
  \mathcal{A}_{s}^{i,j,k,l}(\tau)
  =
  4^{-\frac{1}{3}(h_i+h_j+h_k+h_l)}
  \eta(\tau)^{-2w(i,j,k,l)}
  \mathcal{F}_{il;jk}^{s}(1-\zeta(\tau)) \, .
\end{equation}



\subsection{Dualities and modular covariance}
\label{sec:dualities}

Duality relations act in the space of conformal blocks
\cite{Moore1990}.  The F-duality states the associativity of fusion
product, relating conformal blocks singular near $\zeta=0$ to those
singular near $\zeta=1$:
\begin{equation}
  \label{eq:f-duality}
  \mathcal{F}_{ij;kl}^{s}(\zeta)=\sum_{r}
  F_{sr}{\scriptsize \left[ \begin{array}{cc}
           i & l \\
           j & k \end{array} \right]}
     \mathcal{F}_{il;jk}^{r}(1-\zeta)\, .
\end{equation}
Another transformation is braiding, exchanging the position of $\phi_j$
and $\phi_k$:
\begin{equation}
  \label{eq:B-duality}
  \mathcal{F}_{ij;kl}^{s}(\zeta)=\sum_{r}
  B_{sr}{\scriptsize \left[ \begin{array}{cc}
           i & l \\
           j & k \end{array} \right]}
     \mathcal{F}_{ik;jl}^{r}\left(\frac{1}{\zeta}\right)\, .
\end{equation}
$F$- and $B$-matrices satisfy several identities \cite{Moore1990}.
The relation between $F$-matrices and boundary OPE coefficients
$C_{ijs}^{abc}$ has been pointed out in eq.~\eqref{eq:C_OPE_F}.

Let us remark now that the space of conformal blocks furnishes a
representation of the modular group $\Gamma$ $=$ PSL$(2,\mathbb{Z})$
with the following natural action:
\begin{equation}
  \label{eq:psl_F}
  M\circ \mathcal{F}(\zeta(\tau))= \mathcal{F}(\zeta(M\tau))\, ,
  \quad M\in \Gamma\, .
\end{equation}
Indeed one can relate $\zeta(M\tau)$ to $\zeta(\tau)$ using the
explicit expression of the anharmonic ratio in terms of the modular
parameter, eq.~\eqref{eq:anratio}. For the generators
$S:\tau\to-1/\tau$ and $T:\tau\to\tau +1$, one has
\begin{align}
  \zeta(S\tau)=1-\zeta(\tau)\, ; \quad 
  \zeta(T\tau)=\frac{1}{\zeta(\tau)}\, ,
\end{align}
so that the action of $T$ and $S$ on conformal blocks is
\begin{align}
  S\circ \mathcal{F}_{ij;kl}^{s}(\zeta(\tau))
  &= \sum_{r}
  F_{sr}{\scriptsize \left[ \begin{array}{cc}
           i & l \\
           j & k \end{array} \right]}
     \mathcal{F}_{il;jk}^{r}(\zeta(\tau))\\
  T\circ \mathcal{F}_{ij;kl}^{s}(\zeta(\tau))
  &= 
  \sum_{r}
  B_{sr}{\scriptsize \left[ \begin{array}{cc}
           i & l \\
           j & k \end{array} \right]}
     \mathcal{F}_{ik;jl}^{r}(\zeta(\tau))\, .
\end{align}
This action transfers immediately into modular covariance of
rectangular amplitudes using eq.~\eqref{eq:ish_cb}.  Modular inversion
$S$ acts in an obvious way on the rectangle by interchanging its
length and width. Modular translation $T$ instead deforms it to a
rhombus, and its physical meaning could be understood by using the action
given in the equation above as braiding of the boundary fields.  Let
us see then in more details how $\Gamma$ acts on the rectangle.
$S$-action allows to relate amplitudes of rectangles of inverse aspect
ratio as
\begin{align}
  \label{eq:S_ampli}
  S\circ \mathcal{A}_{s}^{i,j,k,l}(\tau)
  &=
  \mathcal{A}_{s}^{l,i,j,k}(-1/\tau)\\
  &=
  \tau^{-w(i,j,k,l)}
  \sum_r F_{sr}{\scriptsize \left[ \begin{array}{cc}
           i & l \\
           j & k \end{array} \right]}
     \mathcal{A}_{r}^{i,j,k,l}(\tau) \, ,
\end{align}
where we defined
\begin{equation}
  \mathcal{A}_{s}^{l,i,j,k}(-1/\tau)= \bsbra{l}{k}{s}
  \hat{\tilde{q}}^{L_0-c/24} \bsket{i}{j}{s}\, ,
\end{equation}
with 
\begin{equation}
  \label{eq:q_t_tau}
  \hat{\tilde{q}}=\sqrt{\tilde{q}}=\exp(-\pi i /\tau)  \, .
\end{equation}
Note the appearance of the modular weight $w(i,j,k,l)$ (not present
if considering the transformation of the full CFT partition function $Z$,
because of the relation \eqref{eq:Z_ampli2})
from the transformation of $\eta$, see Appendix \ref{sec:usef_form}.
To see how $T$ acts, we consider its action on the amplitude of a
rectangle of inverse aspect ratio
\begin{align}
  \label{eq:T_ampli}
  T\circ 
    \mathcal{A}_s^{i,j,k,l}(-1/\tau) 
  &= \mathcal{A}_s^{i,j,k,l}(-1/(\tau+1))\\ 
  &=e^{i\pi w(i,j,k,l)/3} (\tau+1)^{-w(i,j,k,l)}\eta(\tau)^{-2w(i,j,k,l)} 
  \mathcal{F}^s_{ij;kl}\left(\frac{1}{\zeta(\tau)}\right) \\
  \label{eq:T_cb}
  &  =e^{i\pi w(i,j,k,l)/3}
  \left(\frac{\tau+1}{\tau}\right)^{-2w(i,j,k,l)}
  \sum_{r}
  B_{sr}{\scriptsize \left[ \begin{array}{cc}
           i & l \\
           j & k \end{array} \right]}
     \mathcal{A}^{i,k,j,l}_{r}(-1/\tau)\, .
\end{align}
Note that the amplitude $ \mathcal{A}^{i,k,j,l}_{r}(-1/\tau)$ can be
thought of as the $r$-channel amplitude of a rectangle whose bottom,
where operators $\phi_j^{a,b}$ and $\phi_k^{b,c}$ sit, has been
twisted by $\pi$:
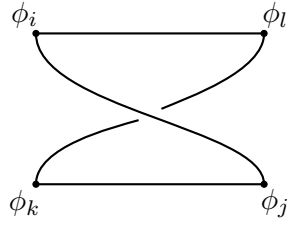
\begin{center}
\begin{tikzpicture}
  \draw[thick] (0,0)--(3,0);
  \draw[thick] (0,-2)--(3,-2);
  \draw[thick] (0,0) .. controls (0,-1) and (3,-1.25) .. (3,-2);
  \draw[thick] (3,0) .. controls (3,-0.5) and (2,-0.85) .. (1.65,-1);
  \draw[thick] (1.35,-1.15) .. controls (1,-1.25) and (0,-1.5) .. (0,-2);
  \draw[fill=black] (0,-2) circle (0.04); 
  \draw[fill=black] (3,-2) circle (0.04); 
  \draw (-0.15,-2.25) node{$\phi_k$};
  \draw (3.15,-2.25) node{$\phi_j$};
  \draw[fill=black] (0,0) circle (0.04); 
  \draw[fill=black] (3,0) circle (0.04); 
  \draw (-0.15,0.25) node{$\phi_i$};
  \draw (3.15,0.25) node{$\phi_l$};
\end{tikzpicture}
\end{center}
Recall that for the torus a Dehn twist by $2\pi$ at fixed time is
exactly $T$; here we have $\hat{q}$ so the twist is the half. This
should be the way to understand modular translation of rectangular
amplitudes.  In particular if the twisted amplitude is equal to the
original one, as happens when $k=j$, we expect to have invariance (in
the sense of equation \eqref{eq:T_ampli}) under modular translations.

More generally modular transformations are powerful tools for
constraining the objects we want to compute, and their usefulness
are well appreciated for the case of a CFT on cylinders and tori
\cite{Cardy1986,Cardy1989}.  Let us see what symmetry constraint one
has on the rectangle\footnote{ The action of the modular group on a
  rectangle has been discussed in \cite{Kleban2003,Diamantis2009} in
  the case of percolation. } and how the basis state we have
introduced solves it.  The natural constraint comes from imposing the
consistency of the partition function upon switching the direction of
imaginary time from vertical to horizontal:
\begin{equation}
  \label{eq:cons_rect}
  \rectamup{b}{d}{c}{a}
  =
  \rectamright{b}{d}{c}{a}
  {}\, .
\end{equation}
From eq.~\eqref{eq:Z_ampli2}, the l.~h.~s.~is 
\begin{equation}
  \label{eq:cons_rect_lhs}
  \rectamup{b}{d}{c}{a}
  =
  L^{w(i,j,k,l)}
  \sum_s C_{ils}^{adc}C_{jks}^{abc}C_{ss0}^{aca}
  \mathcal{A}_{s}^{i,j,k,l}(\tau)\, ,
\end{equation}
while the r.~h.~s.~is
\begin{align}
  \label{eq:cons_rect_rhs}
  \rectamright{b}{d}{c}{a}
  &=
  L'^{w(i,j,k,l)}
  \sum_s C_{ijs}^{dab}C_{kls}^{bcd}C_{ss0}^{dbd}
  \bsbra{k}{l}{s}
  \hat{\tilde{q}}^{L_0-c/24}
  \bsket{i}{j}{s} \\
  &=
  L'^{w(i,j,k,l)}
  4^{-\frac{1}{3}(h_i+h_j+h_k+h_l)}
  \sum_s C_{ijs}^{dab}C_{kls}^{bcd}C_{ss0}^{dbd}
  \eta(-1/\tau)^{-2w(i,j,k,l)}  \\ \nonumber
  &\qquad\qquad\qquad\qquad\qquad\qquad\qquad\qquad\qquad\qquad
  \times \mathcal{F}_{ij;kl}^{s}(1-\zeta(-1/\tau))\\
  &=
  L^{w(i,j,k,l)}
  \sum_{r}\sum_s C_{ijs}^{dab}C_{kls}^{bcd}C_{ss0}^{dbd}
  F_{sr}{\scriptsize \left[ \begin{array}{cc}
        i & l \\
        j & k \end{array} \right]}
  \mathcal{A}_{r}^{i,j,k,l}(\tau)   \, .
\end{align}
Equating the coefficients of the amplitude we get the following
consistency equation:
\begin{equation}
  \label{eq:cons_rect_F}
  \sum_s C_{ijs}^{dab}C_{kls}^{bcd}C_{ss0}^{dbd}
  F_{sr}{\scriptsize \left[ \begin{array}{cc}
        i & l \\
        j & k \end{array} \right]}
  =
  C_{ilr}^{adc}C_{jkr}^{abc}C_{rr0}^{aca}\, .
\end{equation}
This equation is simply stating the associativity of the fusion
product of boundary fields, the so-called sewing constraint coming
from crossing symmetry of correlators~\cite{Lewellen1992}. It has been
shown in \cite{Runkel1999} that its solution is given by
\eqref{eq:C_OPE_F}.  Then, since we defined directly our boundary states
in terms of physical boundary OPEs, consistency under modular
inversion does not give any additional constraint. Recall that for the
cylinder geometry a similar argument gives the so-called Cardy states
\cite{Cardy1989}.


\subsection{Examples}
\label{sec:ex}

We now show how to construct boundary states and associated amplitudes
in some explicit examples. The case of homogeneous boundary
conditions---no BCC operator insertions at the corners---has already
been discussed in details in \cite{Bondesan2011}, and we here
concentrate on the change of boundary conditions.

\subsubsection{Two operator insertions}
\label{sec:two_op}

The simplest case to deal with is that of two different boundary
conditions around the rectangle, when two BCC operator are inserted in
two corners.  This computation amounts to transforming the two-point
function of the BCC operators in the upper hal plane to the rectangle geometry
with operators in the corners and has already appeared in
\cite{Kleban1992}.  In this case only one channel can propagate and
our formulas of section \ref{sec:ampl_cb} are of course simplified
considerably.  For definiteness let  us consider first the case of
boundary condition $a$ on the bottom and $b$ on the other sides,
and call $\phi_i^{a,b}$ the BCC operator. 
The partition function is:
\begin{align}
  \label{eq:Z_two_op_bot}
  \Zrect{a}{b}{b}{b}&=
  \rectamup{a}{b}{b}{b}=
  \Zrect{b}{a}{b}{b}=
  \rectamup{b}{a}{b}{b}  \\
  &= L^{c/4-4 h} \langle B_{bb}^b| \hat{q}^{L_0-c/24}|B_{bb}^a\rangle\\
  &= L^{c/4-4 h} C_{ii0}^{aba} \mathcal{A}_0^{0,i,i,0}(\tau)
  \, .
\end{align}
Using eq.~\eqref{eq:ish_cb}, we express the amplitude using the 
following conformal block:
\begin{equation}
  \label{eq:F_2pt_bot}
  \mathcal{F}_{00;ii}^0(1-\zeta)=\left(\frac{\zeta}{(1-\zeta)^2}\right)^{h/3}
  =4^{-4/3h}\left(\frac{\eta(\tau)}{\eta(2\tau)}\right)^{8h}\, ,
\end{equation}
to have:
\begin{align}
  \Zrect{a}{b}{b}{b}
  &= 
  C_{ii0}^{aba} 4^{-2h}
  L^{c/4-4 h} \eta(\tau)^{-c/2+16 h}\eta(2\tau)^{-8 h}\\
  &= C_{ii0}^{aba} 4^{-2h} L^{c/4-4 h} 
  q^{-c/48}\\
  \label{eq:Z_two_op_bot_series}
  &\quad\times\left[
    1 
    + \frac{c - 32 h}{2} q 
    + \left(\frac{c(6 + c)}{8} - 8 (2 + c) h + 128 h^2\right) q^2  
    + \dots \right] \, .
\end{align}
Note that as expected from the symmetry of interchanging the boundary
operators, this partition function is invariant (up to a phase) with
respect to modular translation. Of course it is not invariant under
modular inversion $S$, and indeed under $S$ this partition function
transforms into that for operators inserted in the left or right corners,
which is:
\begin{align}
  \label{eq:Z_two_op_left}
  \Zrect{b}{b}{b}{a}
  &=
  \rectamup{b}{b}{b}{a}=
  \Zrect{b}{b}{a}{b}
  =
  \rectamup{b}{b}{a}{b}\\
  &= 
  L^{c/4-4 h}   \langle B_{ab}^b| \hat{q}^{L_0-c/24}|B_{ab}^b\rangle \\
  &= 
  L^{c/4-4 h}C_{ii0}^{aba} \mathcal{A}_i^{i,i,0,0}(\tau) \\
  &=C_{ii0}^{aba}  
  L^{c/4-4 h}\eta(\tau)^{-c/2+16 h}\eta(\tau/2)^{-8 h} \\
  \label{eq:Z_two_op_left_series}
  &=C_{ii0}^{aba} 
  L^{c/4-4 h}
  q^{-c/48+h/2}
  \left[
    1 
    + 8 h \hat{q}
    + \frac{c + 8h(8h-1)}{2} q   
    + \dots \right]  \, .
\end{align}
If instead the operators are inserted in non-adjacent corners, say in
the top-left and the right-bottom one, we then have 
\begin{align}
  \label{eq:Z_two_op_lr}
  \Zrect{a}{b}{a}{b}&=\rectamup{a}{b}{a}{b} = \Zrect{b}{a}{a}{b}
  =\rectamup{b}{a}{a}{b}\\
  &=
  L^{c/4-4 h} \langle B_{ba}^b| \hat{q}^{L_0-c/24}|B_{ba}^a\rangle\\
  &=
  L^{c/4-4 h} C_{ii0}^{bab} \mathcal{A}_i^{0,i,0,i}(\tau)\, .
\end{align}
The following conformal block
\begin{equation}
  \label{eq:F_2pt_lr}
  \mathcal{F}_{0i;i0}^i(1-\zeta)=\left(\zeta(1-\zeta)\right)^{h/3}
  =2^{4/3h}\left(\frac{\eta(2\tau)\eta(\tau/2)}{\eta(\tau)^2}\right)^{8h}\, ,
\end{equation}
contributes to the amplitude, and we have:
\begin{align}
  \Zrect{a}{b}{a}{b}
  &= C_{ii0}^{bab} L^{c/4-4 h} \eta(\tau)^{-c/2-8 h}\eta(\tau/2)^{8 h} 
  \eta(2\tau)^{8 h} \\
  \label{eq:Z_two_op_lr_series}
  &= C_{ii0}^{bab} L^{c/4-4 h} 
  q^{-c/48+h/2}
  \left[
    1 
    - 8 h \hat{q}
    + \frac{c + 8 h (8 h - 1)}{2} q   
    + \dots \right]  \, .
\end{align}

The above results can be also obtained by computing the overlap
of the boundary states. 
The amplitude $\mathcal{A}_0^{0,i,i,0}(\tau)$ involved in the partition function
of eq.~\eqref{eq:Z_two_op_bot}, is defined as in eq.~\eqref{eq:ampli_ish},
so it will be given by the overlap of the boundary 
state for homogeneous boundary conditions, see eq.~\eqref{eq:Bhom},
and $\bsket{i}{i}{0}$, whose first levels read:
\begin{equation}
  \label{eq:Bbba_id}
  \begin{split}
   4^{2h} \bsket{i}{i}{0}=
  |0\rangle
  +
  \left(\frac{32 h}{c}-1\right) L_{-2}|0\rangle
  +
  \frac{c (22 + 5 c) - 64 (6 + 5 c) h + 5120 h^2}{2 c (22 + 5 c)}
  L_{-2}^2|0\rangle\\
  -
  \frac{c (22 + 5 c) - 256 (2 + c) h + 3072 h^2}{2 c (22 + 5 c)} 
  L_{-4}|0\rangle
  +\dots 
  \end{split}
\end{equation}
We have checked that $\langle B_{bb}^{b}| \hat{q}^{L_0-c/24}
|B_{bb}^{a}\rangle$ computed using the above expression of boundary
states reproduces the small-$q$ expansion of
eq.~\eqref{eq:Z_two_op_bot_series} up to order $\hat{q}^{10}$.
The boundary states building the amplitudes of
eq.~\eqref{eq:Z_two_op_left} and \eqref{eq:Z_two_op_lr} are given by
a single basis state as in \eqref{eq:Baac}. 
Eq.~\eqref{eq:Z_two_op_left} is given by the overlap of:
\begin{equation}
  \label{eq:Baab_first_lev}
  \begin{split}
    \bsket{i}{0}{i}
    =
    |\phi_i\rangle
    -2 L_{-1}|\phi_i\rangle
    -L_{-2}|\phi_i\rangle  + 2 L_{-1}^2|\phi_i\rangle
    +2 L_{-2}L_{-1}|\phi_i\rangle 
    - \frac{4}{3} L{-1}^3|\phi_i\rangle\\ 
    -\frac{1}{2} L_{-4}|\phi_i\rangle  
    + \frac{1}{2} L_{-2}^2 |\phi_i\rangle  
    - 2 L_{-2}L_{-1}^2 |\phi_i\rangle  
    + \frac{2}{3} L_{-1}^3|\phi_i\rangle  
    +\dots
  \end{split}
\end{equation}
with itself.  Instead the partition function in
eq.~\eqref{eq:Z_two_op_lr} is given by $\bsbra{0}{i}{i}
\hat{q}^{L_0-c/24}\bsket{i}{0}{i}$.  $\bsbra{0}{i}{i}$ is computed
from $\bsbra{i}{0}{i}$ by sending $2 \to -2$ in the explicit formula
\eqref{eq:Baac} and this gives the minus signs present in equation
\eqref{eq:Z_two_op_lr_series} with respect to
\eqref{eq:Z_two_op_left_series}.  Again we have verified the agreement
of the series expansions using the explicit form of the boundary
states up to level $\hat{q}^{10}$. Note that the small-$q$ expansion
\eqref{eq:Z_two_op_bot_series} contains only even powers of $\hat{q}$,
as expected from left-right symmetry of the boundary conditions.
Indeed odd powers are related to
descendants at odd levels changing sign under inversion of the
coordinates $z\to -z$, and thus are not allowed to propagate.

Furthermore one can check the agreement of these results with
computations for free theories presented in section
\ref{sec:free_theo},
eqs.~\eqref{eq:DDDN}, \eqref{eq:NDDN} and \eqref{eq:DNNN}.

\subsubsection{Four operator insertions}
\label{sec:four_op}

We now turn to the case of four insertions of BCC operators, and we
consider the simple situation when the operators inserted are all
degenerate at level $2$, so that the dimension of each operator equals
$h_{1,2}$ or $h_{2,1}$.  We make this choice since an explicit and
simple expression for the conformal blocks of these fields is available
\cite{DiFrancesco1997} (actually it suffices that just one of the four
fields is degenerate at level $2$ to find a closed form of the
conformal blocks, but we treat the case of all fields degenerate at
level $2$ for easiness of notation).  For fixing ideas we can think of
the $Q$-states Potts model with fixed boundary conditions on left and
right boundaries and free on the top and bottom sides. Clearly we have
to distinguish two situations: either the fixed spins are in an equal
state $\alpha$ or they are different, say in states $\alpha$ and
$\alpha'$. These partition functions have
been already discussed in details in
\cite{Cardy2001} and here we reproduce the
results and comment about boundary states and modular properties. 
Introducing the label $i$ of the
Virasoro modules $V_{(1,1+i)}$ (or $V_{(1+i,1)}$) in Kac notation, one
associates to free boundary conditions on the strip the label $1$, to
fixed equal $0$, and to fixed different $0$ and $2$. This is
consistent with the known fusion rules \cite{Cardy2001} since
$1\otimes 1=0\oplus 2$.  Then concretely we would like to compute the
partition functions
\begin{equation}
  \label{eq:def_Z00_20}
  \Zrect{1}{1}{a}{0}= 
  \rectamup{1}{1}{a}{0}
  \, , \qquad a = 0,2\, .
\end{equation}
If $a=0$, at the corners on the bottom there are BCC operators
$\phi_{1}^{0,1}$ and $\phi_{1}^{1,0}$ whose OPE coefficients are
clearly $C_{1,1,s}^{0,1,0}\propto\delta_{s,0}$ so that only the
identity channel propagates. For $a=2$ we insert instead
$\phi_{1}^{0,1}$ and $\phi_{1}^{1,2}$ whose OPE coefficients are
$C_{1,1,s}^{0,1,2}\propto\delta_{s,2}$ so that only the
$\phi_{2}$-channel is left. Calling $h\equiv h_{1}$ and
$\mathcal{A}_a\equiv\mathcal{A}_a^{1,1,1,1}$ , we then have
\begin{equation}
  \label{eq:Z_Am_00_02}
  \Zrect{1}{1}{a}{0}= \tilde{\mathcal{N}}^a
  L^{c/4-8h}\mathcal{A}_a\, .
\end{equation}
The normalization $\tilde{\mathcal{N}}^a$ can be written in terms
of OPE constants, and its value will be fixed later.
We turn then our attention to the computation of the conformal blocks
$\mathcal{F}_{1,1;1,1}^{a}$ present in the expression of the
amplitudes of eq.~\eqref{eq:ish_cb}. This is a standard exercise 
\cite{DiFrancesco1997}. 
From the null state condition at level $2$, the function $\mathcal{G}^{a}$,
\begin{equation}
  \label{eq:G_def}
  \mathcal{G}^{a}(\zeta)
  :=
  \left( \zeta (1-\zeta) \right)^{4h/3}
  \mathcal{F}_{1,1;1,1}^{a}(\zeta)\, ,
\end{equation}
satisfies the hypergeometric equation
\begin{equation}
  \zeta(1-\zeta)\mathcal{G}''+
  \left\{ \gamma -(1+\alpha +\beta )\zeta \right\}\mathcal{G}'-
  \alpha \beta \mathcal{G}=0\, .
\end{equation}
with parameters
\begin{equation}
  \label{eq:hyp_par}
  \alpha = -4h, \beta =
  \frac{1}{3}-\frac{4}{3}h, \gamma = \frac{2}{3}-\frac{8}{3}h\, .  
\end{equation}
The roots of the indicial equation 
\begin{equation}
  \label{eq:roots_ind}
  r_1=0\, , \quad r_2=1-\gamma\, ,
\end{equation}
label the two solutions of the
differential equation (we take the $\sim 0$ solutions), defining the
conformal blocks of the identity $\phi_0$, $\mathcal{G}^{0}(\zeta)$,
and that of the operator $\phi_2$ of weight $h_2=8h/3+1/3$,
$\mathcal{G}^{2}(\zeta)$:
\begin{align}
  \label{eq:conf_block_0_phi12}
  \mathcal{G}^{0}(\zeta) &= \, _2F_1\left(\frac{1}{3}-\frac{4 h}{3},-4
    h;\frac{2}{3}-\frac{8 h}{3};\zeta\right)\\
  \label{eq:conf_block_1_phi12}
  \mathcal{G}^{2}(\zeta)&=
  \zeta^{\frac{8 h}{3}+\frac{1}{3}} \, _2F_1\left(\frac{1}{3}-\frac{4 h}{3},\frac{4
      h}{3}+\frac{2}{3};\frac{8 h}{3}+\frac{4}{3};\zeta\right)\, .
\end{align}
The validity of these solutions is for $\gamma$ not an
integer.

After a little algebra the partition function  reads
\begin{equation}
  \label{eq:Zfinal_N}
  \Zrect{1}{1}{a}{0}
  =\mathcal{N}^a
  L^{c/4-8h}
  \eta^{-c/2} \theta_3^{16 h} \mathcal{G}^a(1-\zeta)\, ,
\end{equation}
where we reabsorbed all constants in the definition of
$\mathcal{N}^a$. The relevant $F$-matrices for the conformal blocks
involved can be computed using standard hypergeometric
identities relating the conformal blocks of the correlator $\langle
\phi_{1}\phi_i\phi_j\phi_k\rangle$, allowing explicit checks that they
behave as expected under modular inversion.
We report here for future convenience the $F$-matrices 
$F_{ij}:=
F_{ij}{\scriptsize \left[ \begin{array}{cc}
           1/2 & 1/2 \\
           1/2 & 1/2 \end{array} \right]}$:
\begin{equation}
  \label{eq:Fij}
  \begin{pmatrix}
    F_{00} & F_{02}\\
    F_{20} & F_{22}\\
  \end{pmatrix}
  =
  \begin{pmatrix}
    \smfrac{1}{\beta} & \smfrac{\Gamma \left(-\smfrac{8 h}{3}-\smfrac{1}{3}\right) \Gamma \left(\smfrac{2}{3}-\smfrac{8
   h}{3}\right)}{\Gamma \left(\smfrac{1}{3}-\smfrac{4 h}{3}\right) \Gamma (-4 h)}\\
    \smfrac{\Gamma \left(\smfrac{8 h}{3}+\smfrac{1}{3}\right) \Gamma \left(\smfrac{8
   h}{3}+\smfrac{4}{3}\right)}{\Gamma \left(\smfrac{4 h}{3}+\smfrac{2}{3}\right) \Gamma (4 h+1)} & -\smfrac{1}{\beta}\\
  \end{pmatrix}\, ,
\end{equation}
where we defined $\beta=2 \sin\left(\pi (8h+1 )/6\right)
=2\cos(\pi/(p+1))$, 
if we parametrize as usual $c=1-6/(p(p+1))$, and $h=h_{1,2}$.
$\beta$  will play the role
of the weight of loops in the study of loop models in the next sections.

A small-$q$ expansion of the amplitudes $\mathcal{A}_a$ can also be
computed using the explicit expression of the associated boundary
states (see Appendix \ref{sec:bs_two_one_leg}). To first order we have
\begin{align}
  \label{eq:smallq-Z00}
  \mathcal{A}^0(\tau)
  &=4^{-4h_{1}}
  q^{-c/48}
  \left(
    1
    +
    \frac{(3 - 7 p)^2 (-2 + p)}{2 p (1 + p) (3 + p)}q
    +\dots \right)
  \\
  \label{eq:smallq-Z01}
  \mathcal{A}^1(\tau)
  &=4^{2h_2-4h_{1}}
  q^{-c/48+h_2 /2}
  \left(
    1
    +
    \left( 
      \frac{9}{2} - \frac{3}{p} - \frac{45}{1 + p} + \frac{80}{1 + 3 p}
      \right) q
    +\dots \right)\, .
\end{align}
We checked agreement with the expansion of eq.~\eqref{eq:Zfinal_N} up to
order $\hat{q}^{10}$.

Finally, to interpret the partition function above as that of the Potts
model, we fix the normalization following
\cite{Cardy2001}, demanding that in the limit $\zeta\to 0$ of an
infinitely long rectangle, the normalized conformal
blocks\footnote{
Redefining $\phi_i^{ab}\to \lambda_i^{ab} \phi_i^{ab}$
  and $|0\rangle\to \alpha |0\rangle$ we can
  fix opportunely say $\lambda_{1/2}^{0,1/2}$ 
  and $\lambda_{1/2}^{1,1/2}$ so that the OPE structure constants
  give the behavior of
  eq.~\eqref{eq:Ga_norm}.} behave as
\begin{equation}
  \label{eq:Ga_norm}
  \mathcal{N}^a \mathcal{G}^a(1-\zeta)= 1 + O(\zeta)\, .  
\end{equation}
This holds because for a very long rectangle, one has in both cases 
$a=0$, $a=2$, 
the partition function of an horizontal strip with free boundary
conditions on the sides. Both conformal blocks
should give one in that limit. 
We have:
\begin{align}
  \label{eq:N0}
  \mathcal{N}^0 &=\frac{1}{F_{00}} 
  = \beta\\
  \label{eq:N1}
  \mathcal{N}^2 &= \frac{1}{F_{20}}=  \frac{\Gamma \left(\frac{4h}{3}+\frac{2}{3}\right) 
    \Gamma (4h+1)}
  {\Gamma
   \left(\frac{8 h}{3}+\frac{1}{3}\right) 
   \Gamma \left(\frac{8h}{3}+\frac{4}{3}\right)} \, .
\end{align}

\section{The case of free theories}
\label{sec:free_theo}

We focus now on the description of rectangular amplitudes in free
theories. This section completes the discussion of free theories
presented in \cite{Bondesan2011} with examples with different boundary
conditions along the rectangle.

\subsection{The Laplacian of the rectangle}
\label{sec:lapl}

We start our discussion of 
free theories 
by computing the partition function of (unrooted) spanning trees on a
rectangle.  This problem is described by a free fermionic field theory via
the matrix-tree theorem, stating that the number of spanning trees on
a graph $G$ is given by the determinant of a minor of the Laplacian
matrix $\Delta$ of G (see e.g.~the review \cite{Sokal2005}):
\begin{equation}
  \label{eq:ztree}
  \Ztree_G = \det \left(\Delta_{\backslash i} \right)\, .
\end{equation}
where $\Delta_{\backslash i}$ is the matrix obtained from $\Delta$ by
deleting the $i$th row and column, to eliminate the zero eigenvalue.
Then the number of spanning trees is given by a two point
function in the fermionic theory.
Neumann boundary conditions ($N$) for the field translate to free for the
spanning trees, while Dirichlet ones ($D$) impose that boundary sites are
linked to a single external vertex (``wired'' boundary conditions).

We would like now to compute the Laplacian with either $N$ or $D$ on the
sides of a rectangle and take the continuum limit. This can be
done by taking the tensor product of Laplacians on a
segment, and from that one can extract the expressions for the
universal part in the limit of infinite rectangle with fixed
$\tau=i L'/L$.  Some of these results were first given in \cite{David1988}
and one finds: 
\begin{align}
  \label{eq:NNNN}
  \rectamup{D}{D}{D}{D} &= \rectamup{N}{N}{N}{N} =
  \frac{\eta(\tau)}{\sqrt{L}}\\
  \label{eq:NNDD}
  \rectamup{N}{N}{D}{D} &\propto \sqrt{L}\eta(\tau)\\
  \label{eq:DDDN}
  \rectamup{D}{D}{D}{N} &= \rectamup{N}{N}{N}{D} \propto
  \frac{\eta(\tau/2)}{\eta(\tau)}\\
  \label{eq:NDDN}
  \rectamup{N}{D}{D}{N} &\propto
  \frac{\eta^2(\tau)}{\eta(\tau/2)\eta(2\tau)}\, .
\end{align}
In these results we did not keep track of proportionality
coefficients, which are related to the OPE structure constants, and in
principle different in each case\footnote{
Moreover the free energies computed
from the lattice Laplacian contain also an additional
geometry-independent constant, which we cannot predict using CFT.
For example in the case of $D$ boundary conditions around the rectangle
it is \cite{David1988} $F_0 = \smfrac{1}{2}\log(4\sqrt{2})$.
For a recent discussion of this issue in the case of the Ising
model with free boundary conditions, see \cite{Wu2012}.}.
 The partition functions above are
enough to determine all the other obtained by permuting boundary
conditions.  This is possible using the symmetry of describing the
same partition function choosing imaginary time flowing either bottom-top
or left-right (crossing symmetry of correlators of boundary fields),
and modular covariance of these partition functions.  For example from
modular inversion one has:
\begin{align}
  \label{eq:DNNN}
  \rectamup{D}{N}{N}{N} 
  \propto \frac{\eta(2\tau)}{\eta(\tau)} \, ,
\end{align}
where we have used identities listed in Appendix \ref{sec:usef_form}.
Note in particular that the partition functions \eqref{eq:NNNN} and
\eqref{eq:NDDN} are modular invariant.  
The most important physical feature exhibited by
\eqref{eq:NNNN}-\eqref{eq:NDDN} is the dimension-full factor
$L^{w}$, with $w$ the modular weight given by eq.~\eqref{eq:mod_weight}.
Indeed in the continuum the fermionic theory describing
spanning trees is the logarithmic CFT of symplectic
fermions with $c=-2$, where the twist field changing boundary
conditions $N$ to $D$ and vice versa, has conformal dimension
$h_{1,2}=-1/8$ \cite{Ivashkevich1999}.

Having discussed the case of $c=-2$ one can immediately find the
result for $c=1$, the free boson, by taking the power $-1/2$ of the
results above since the twist field in this case has dimension
$1/16=(-1/2)(-1/8)$. 
The case of a free boson on a rectangle was discussed in detail in
\cite{Imamura2006}, and our results are in agreement with those
findings. Further, in that work the authors discussed also the
boundary states associated to different choices of boundary
conditions. The results resemble those for the cylinder boundary
states except that, as expected from the general discussion of section
\ref{sec:sol_gluing}, for the rectangle only one set of oscillator modes
$a_n$ of the bosonic field is left.  The boundary state describing the
bottom of the rectangle is \cite{Imamura2006}
\begin{equation}
  \label{eq:Bstate_bos}
  \left| B^{\epsilon_b}_{\epsilon_l,\epsilon_r} \right>
  =
  \exp\left(-\epsilon_l\epsilon_b\sum_{n>0}\frac{1}{2n}a_n^2 \right)
  \left|\theta^{\epsilon_b}_{\epsilon_l,\epsilon_r}\right>\, ,
\end{equation}
where $\epsilon_{l,r,b}$ equal to $+1$ ($-1$) means $N$
($D$) boundary conditions at left, right and bottom side
respectively, and $\left|\theta^{\epsilon_b}_{\epsilon_l,\epsilon_r}\right>$
is the zero mode, which is the simple Fock vacuum except when 
$\epsilon_l=\epsilon_r=+1$, in which case is $\left|p=0\right>$
for $\epsilon_b=1$ and $\left|x=x_0\right>$ for $\epsilon_b=-1$.


\subsection{The Ising model}
\label{sec:ising}

In \cite{Bondesan2011,Imamura2007} the case of the Ising model with $D$
boundary conditions on all sides of the rectangle was considered,
corresponding to fixing Ising spins to the same state on the boundary.
 The
result for the corresponding rectangle
boundary state has the following coherent
state form
\begin{equation}
\label{eq:majo_coherent}
|B_{\psi}^o\rangle
= 
\exp\left(
\sum_{0\leq m< n}^\infty G_{m,n}\psi_{-m-1/2}\psi_{-n-1/2}
\right)|O\rangle \, ,
\end{equation}
with $|O\rangle$ the vacuum annihilated by the positive modes of
the Neveu-Schwartz fermions, $\psi_{p+1/2}|O\rangle=0, 
p\in \mathbb{Z}_{\ge 0}$, and $G_{m,n}$ defined by
\begin{equation}
  G(z_1,z_2)=
  \frac{1}{2(z_2-z_1)}
  \left(\frac{\sqrt{1-\frac{1}{z_1^2}}\sqrt{1-\frac{1}{z_2^2}}}
    {1-\frac{1}{z_1z_2}}-1\right)
  = \frac{1}{2z_1z_2}\sum_{m,n=0}^\infty \frac{G_{mn}}{z_1^{m}z_2^{n}}\, .
\end{equation}

Here we will generalize this result to the case of different boundary
conditions on the sides of the rectangle. This will be done by mapping
correlators from the region with corners to the upper half plane,
through the conformal map $f(z)=(z+z^{-1})/2$, similar
to the situation of figure \ref{fig:map_D_UHP}.

Let us consider first the upper half plane $\mathbb{H}$
with boundary conditions for Ising spins free $f$ on $\{x<1\}\cup \{x>1\}$
and fixed (say $+$) in between.
This corresponds to inserting a BCC operator $\sigma^{f,+}$ at
$x=\pm 1$. 
From the fusion rules of these BCC operators 
\begin{equation}
\sigma^{f,+} \otimes \sigma^{+,f} =I^{f,f}+\psi^{f,f}\, ,
\end{equation}
we know that the corresponding rectangle boundary state $\BffF$ is
composed by two basis states, one in the sector of the identity
$\bsket{\sigma}{\sigma}{I}$ and one in that of the energy
$\bsket{\sigma}{\sigma}{\psi}$:
\begin{equation}
  |B_{ff}^+\rangle
  \propto \bsket{\sigma}{\sigma}{I} + C \bsket{\sigma}{\sigma}{\psi}\, .
\end{equation}
The numerical value of $C$ is
\begin{equation}
  \label{eq:Ctheo}
  C=\frac{C_{\sigma,\sigma,\psi}^{f,+,f}\sqrt{C_{\sigma,\sigma,I}^{f,f,f}}}
  {C_{\sigma,\sigma,I}^{f,+,f}\sqrt{C_{I,I,I}^{f,f,f}}}
  =4^{-h_{\psi}}\sqrt{2}  \, ,
\end{equation}
as follows from using $F$-matrices for the Ising model (see
e.~g.~\cite{Lewellen1992}) after identification of boundary conditions
with Virasoro representations $f\equiv \sigma$, $+\equiv I$ and
normalizing two point functions to one.

In order to evaluate the first contribution $\bsket{\sigma}{\sigma}{I}$
to the boundary state $|B_{ff}^+\rangle$  we 
consider the two point correlation function of fermions in the upper
half plane with these boundary conditions  (see
e.~g.~\cite{Ardonne2010}, eq.~(22))
\begin{equation}
\label{eq:psi_H}
\begin{split}
\langle \psi(w_1)\psi(w_2)\rangle_{f+f,\mathbb{H}}
=
{1\over 2(w_{1}-w_2)}\Big[\left({1+w_1\over 1+w_2}\right)^{1/2}
\left({1-w_2\over 1-w_1}\right)^{1/2}+\\
\left({1+w_2\over 1+w_1}\right)^{1/2} \left({1-w_1\over 1-w_2}\right)^{1/2}\Big]
\end{split}
\end{equation}
where we obtained this result from the chiral
correlator $\langle \psi(w_1)\psi(w_2)\sigma(1)\sigma(-1)\rangle$ and
omitting irrelevant factors.  This result can be related to the
boundary state $\BffF$ describing the geometry ${\cal D}$ with $f+f$
boundary conditions through
\begin{align}
\label{eq:corr_D_H}
\langle \psi(z_1)\psi(z_2)\rangle_{f+f,{\cal D}}
= \langle O|\psi(z_1)\psi(z_2)\BffF
=\left(\frac{\partial w_1}{\partial z_1}\frac{\partial w_2}{\partial z_2}
\right)^{1/2}\langle\psi(w_1)\psi(w_2)\rangle_{f+f,\mathbb{H}}\, .
\end{align}
Following the discussion in \cite{Bondesan2011}, we look for 
a fermionic coherent state form of the boundary state
\begin{equation}
\bsket{\sigma}{\sigma}{I}=4^{-2h_\sigma}
:\exp\left(\oint \frac{dz}{2i\pi}\oint \frac{dz'}{2i\pi}
\psi(z)G(z,z')\psi(z')\right):|O\rangle\, ,
\end{equation}
where the factor $4^{-2h_\sigma}$ is introduced in order to obtain consistency with the definition
of basis states \eqref{eq:ish_state}.
Substituting this in equation \eqref{eq:corr_D_H} one can easily
determine the function $G(z,z')$.  It is convenient to split
expression \eqref{eq:psi_H} into the sum of two terms. Going through
the calculations leads to the introduction of two functions
\begin{equation}
G^{(1)}={1\over 4(z_2-z_1)}\left[
{\sqrt{1-{1\over z_1^2}}\sqrt{1-{1\over z_2^2}}\over 1-{1\over z_1z_2}}~{(1+{1\over z_1})(1-{1\over z_2})\over (1-{1\over z_1})(1+{1\over z_2})}-1\right]
\end{equation}
and 
\begin{equation}
G^{(2)}={1\over 4(z_2-z_1)}\left[
{\sqrt{1-{1\over z_1^2}}\sqrt{1-{1\over z_2^2}}\over 1-{1\over z_1z_2}}~{(1-{1\over z_1})(1+{1\over z_2})\over (1+{1\over z_1})(1-{1\over z_2})}-1\right]
\end{equation}
for which we define the expansions
\begin{equation}
G^{(i)}={1\over 2z_1z_2}\sum_{m,n=0}^\infty {G_{mn}^{(i)} \over z_1^mz_2^n}
\end{equation}
so that the boundary state gathers the contribution
\begin{equation}
\bsket{\sigma}{\sigma}{I} =4^{-2h_\sigma} \exp\left(\sum_{0\leq m< n}^\infty\left[G_{mn}\right]\psi_{-m-1/2}\psi_{-n-1/2}
\right)|O\rangle
\end{equation}
where we defined $G = G^{(1)}+G^{(2)}$. The first values of $G_{mn}$ read
\begin{eqnarray}
G_{01}=-{3\over 2}\nonumber\\
G_{03}=-{7\over 8},G_{12}=-{3\over 8}\nonumber\\
G_{05}=-{11\over 16},G_{14}=-{1\over 16},G_{23}=-{7\over 8}\nonumber\\
G_{07}=-{75\over 128},G_{16}=-{3\over 128},G_{25}=-{55\over 128},G_{34}=-{63\over 128}\, .
\end{eqnarray}
This leads to the first few terms in the expansion of the boundary state
\begin{equation}
  \bsket{\sigma}{\sigma}{I} =4^{-2h_\sigma}\left[1+{3\over 2}\psi_{-1/2}\psi_{-3/2}+
    {7\over 8}\psi_{-1/2}\psi_{-7/2}+
    {3\over 8}\psi_{-3/2}\psi_{-5/2}+\ldots\right]|O\rangle \,.
\end{equation}
We now determine the other contribution $\bsket{\sigma}{\sigma}{\psi}$ to $\BffF$ in
addition to $\bsket{\sigma}{\sigma}{I}$.  For doing that we use now the  chiral
correlator $\langle
\psi(w_0)\psi(w_1)\psi(w_2)\sigma(1)\sigma(-1)\rangle$ in
\cite{Ardonne2010} (eq. (66) of that reference) and send $w_0$ to
infinity to obtain
\begin{equation}
\langle\psi(\infty)\psi(w_1)\psi(w_2)\rangle_{f+f,\mathbb{H}}
={w_1^2+w_2^2-w_1w_2-1\over (w_1-w_2)(w_1^2-1)^{1/2}(w_2^2-1)^{1/2}}\, .
\end{equation}
Setting $u={1\over z_1}$ and $v={1\over z_2}$, we find the
corresponding correlator in the $z$ plane
\begin{equation}
  \begin{split}
\langle\psi(\infty)\psi(z_1)\psi(z_2)\rangle_{f+f,{\cal D}}
&={(v-u)(1-uv)\over (1-u^2)^{1/2}(1-v^2)^{1/2}}\\
&\quad +{(1-u^2)^{1/2}(1-v^2)^{1/2}\over (v-u)(1-uv)}
\left[v^2 {1-u^2\over 1-v^2}+u^2{1-v^2\over 1-u^2}\right]\, .
  \end{split}
\end{equation}
We represent this expression alternatively as 
\begin{equation}
\langle \psi(\infty) \psi(z_1)\psi(z_2) :\exp\left(\oint {dz\over 2i\pi}\oint {dz'\over 2i\pi}
\psi(z)G(z,z')\psi(z')\right):\psi(0)|O\rangle\, .
\end{equation}
This gives us the lengthy expression
\begin{equation}
  \begin{split}
G(z_1,z_2)&={1\over 2}{(v-u)(1-uv)\over (1-u^2)^{1/2}(1-v^2)^{1/2}}+{1\over 2}{(1-u^2)^{1/2}(1-v^2)^{1/2}\over (v-u)(1-uv)}
\left[v^2 {1-u^2\over 1-v^2}+u^2{1-v^2\over 1-u^2}\right]\\
&\quad -{u^2+v^2-uv\over(u-v)}\, ,
  \end{split}
\end{equation}
where we used that
\begin{equation}
\langle \psi(\infty)\psi(z_1)\psi(z_2)\psi(0)\rangle={u^2+v^2-uv\over v-u}\, .
\end{equation}
We finally expand as before
\begin{equation}
G(z_1,z_2)=\frac{uv}{2}\sum_{m,n\geq 0} G_{mn}u^mv^n\, ,
\end{equation}
from which we get the final result
\begin{equation}
\bsket{\sigma}{\sigma}{\psi}
=4^{h_\psi-2h_\sigma}
\exp\left[\sum_{0\leq m<n} G_{mn}\psi_{-m+1/2}\psi_{-n+1/2}\right]
\psi(0)|O\rangle\, .
\end{equation}
%
In the end one gets the following expression
\begin{eqnarray}
\bsket{\sigma}{\sigma}{\psi}=4^{h_\psi-2h_\sigma}
\left[1-{1\over 2}\psi_{1/2}\psi_{-5/2}-{3\over 8}\psi_{1/2}\psi_{-9/2}-{5\over 16}\psi_{1/2}\psi_{-13/2}-\right.\nonumber\\
\left.{9\over 8}\psi_{-3/2}\psi_{-5/2}-{5\over 8}\psi_{-3/2}\psi_{-9/2}
+\ldots\right]\psi_{-1/2}|O\rangle\, ,
\end{eqnarray}
for the second (odd) contribution to the boundary state. 


Taking overlap of these boundary states we can compute partition
functions with different boundary conditions of Ising spins, and their
agreement with four point functions of BCC operators along the general
scheme of section \ref{sec:ampl_cb}, can be readily checked.  We will
not present here the details of this check, but support our results
with lattice computations for the Ising model instead.

\subsubsection{Lattice Ising model}
\label{sec:lat_ising}

In this section we will compute the scalar products of the discretization
of the boundary state $\BffF$ with excited states in the Ising Hamiltonian.
We consider the Hamiltonian limit of the $2$D critical
Ising model on a strip with free/free boundary conditions:
\begin{equation}
  \label{eq:H_ising}
  H 
  = 
  -\frac{1}{2} 
  \left( 
  \sum_{i=1}^L \sigma_i^z
  +
  \sum_{i=1}^{L-1} \sigma_i^x \sigma_{i+1}^x
  \right) \, .
\end{equation}
Now recall that the Hamiltonian \eqref{eq:H_ising} is related to the
$2$D Ising model in the $\sigma^x$ basis.  Then to make the contact
with the $2$D model, we should rotate $90$ degrees clockwise the spins
of the chain in the $x-z$ plane, that is, we define the ``fixed''
boundary state $\BffF_{L}$ as
\begin{align}
  \label{eq:fixN}
  \BffF_{L}
  &= 
  \ket{\rightarrow \cdots \rightarrow} = \frac{1}{\sqrt{2^{L}}}
  \sum_{\{\mu_i^z = \uparrow,\downarrow \}} \ket{\mu_1^z \dots \mu_L^z} 
  \, .
\end{align}

We diagonalize the chain \cite{Lieb1961} to obtain the 
eigenvectors $|k\rangle_L$, and compute numerically the scalar
products ${}_L\langle B_{ff}^+ |k\rangle_L$. This task is simplified
by using the free fermionic nature of the problem, in a way similar to what we did 
 for  free boundary conditions in \cite{Bondesan2011}.
This time, however, the computations are more complex and we have
restricted the analysis to small system sizes (up to $L=30$ sites).  Before
presenting the results, we recall that due to the anomaly at the
corners (see eq.~\eqref{eq:mod_weight}), scalar products between a
lattice rectangular boundary state $|B\rangle_L$ and the
(normalized) $k$-th excited state $|k\rangle_L$ of the strip Hamiltonian,
are expected to scale as \cite{Bondesan2011}:
\begin{equation}
  \label{eq:mlogBk}
  -\log( {}_L\langle B | k \rangle_L )
  = 
  a_0 L
  +
  a_1 \log L
  +
  a_2
  +
  \frac{a_3}{L}
  +
  \frac{a_4}{L^2}
  +
  O\left(\frac{1}{L^3}\right) \, ,
\end{equation}
with
\begin{equation}
  \label{eq:a1}
  a_1 = 2(h_l+h_r)-\frac{c}{8}\, .
\end{equation}
The value of the CFT scalar product $\langle B | k \rangle$ is then
extracted from 
\begin{equation}
  \label{eq:gamma}
  a_2 = \gamma - \log( \langle B | k \rangle ) \, ,
\end{equation}
$\gamma$ determined normalizing $\langle B | 0 \rangle = 1$. See
section \ref{sec:num} below for a systematic study of these coefficients. For
computing $\langle B|k\rangle$, we actually fit
\begin{equation}
  \label{eq:eq:mlogBk_div}
  -\log\left( \frac{{}_L\langle B | k \rangle_L}{{}_L\langle B|0\rangle_L}
    \right)
  =
  a_2 +
  \frac{a_3}{L}
  +
  \frac{a_4}{L^2}
  +
  \frac{a_5}{L^3}\, .
\end{equation}

We present the data obtained in table \ref{tab:scal_prod_fix_scal}.
The value $a_1=2\times 2/16 - 1/16 = 3/16$ is found with good accuracy
only for the first lower excited states.  The continuum limit of the
excited states of the chain is associated to a CFT state obtained by
the action of negative fermionic modes on the vacuum.  It is important
to distinguish the two sectors, even and odd number of fermions,
corresponding respectively to descendants of the identity $I$ and of
the energy $\psi$ in the CFT.  We replace $\gamma$ in \eqref{eq:gamma}
by $\gamma + \tilde{\gamma}$ if $\ket{k}$ is a descendant of
$\psi$. $\tilde{\gamma}$ is written as minus the logarithm of a
prefactor $\tilde{C}$ multiplying the states in the sector of
$\psi$. Looking at $\langle B_{ff}^+|1\rangle$
(where $|1\rangle=\psi_{-1/2}|0\rangle$) determines
\begin{equation}
  \label{eq:C}
  \tilde{C} =  1.41422 \pm 0.00002\, . 
\end{equation}
This agrees very well with the prediction from the general theory
eq.~\eqref{eq:Ctheo} (the factor $4^{-h_\psi}$ simplifies with
$4^{h_\psi-2h_\sigma}/4^{-2h_\sigma}$ from the normalization of the
basis states $\bsket{\sigma}{\sigma}{\psi}$ and
$\bsket{\sigma}{\sigma}{I}$), providing a lattice measurement of
boundary OPE coefficients.  In table \ref{tab:scal_prod_fix_scal} we
list the scalar products obtained.  The agreement with the CFT prediction
is good, validating
our analytical computation.


\begin{table}[h!c]
  \centering
  \begin{tabular}{|c|c|c|c|}
    \cline{2-4}
    \multicolumn{1}{r|}{} 
    & $h_k$ & numerics & CFT \\
    \hline
    $\langle B_{ff}^+|0\rangle$ & $0$ & $1$ & $1$ \\
    \hline
    $\langle B_{ff}^+|1\rangle$ & $1/2$ & $1$ & $1$ \\
    \hline
    $\langle B_{ff}^+|2\rangle$ & $3/2$ & $0$ & $0$ \\
    \hline
    $\langle B_{ff}^+|3\rangle$ & $2$ & $1.499999\pm 0.000020$ & $3/2=1.5$ \\
    \hline
    $\langle B_{ff}^+|4\rangle$ & $5/2$ & $0.500079\pm 0.000012$ & $1/2=0.5$ \\
    \hline
    $\langle B_{ff}^+|5\rangle$ & $3$ & $0$ & $0$ \\
    \hline
    $\langle B_{ff}^+|6\rangle$ & $7/2$ & $0$ & $0$ \\
    \hline
    $\langle B_{ff}^+|7\rangle$ & $4$ & $0.875178 \pm 0.000020$ & $7/8=0.875$ \\
    \hline
    $\langle B_{ff}^+|8\rangle$ & $4$ & $0.374991 \pm 0.000005$ & $3/8=0.375$ \\
    \hline
    $\langle B_{ff}^+|9\rangle$ & $9/2$ & $0.375315\pm 0.000031$ & $3/8=0.375$ \\
    \hline
    $\langle B_{ff}^+|10\rangle$ & $9/2$ & $1.125002\pm 0.000020$ & $9/8=1.125$ \\
    \hline
  \end{tabular}
  \caption{Scalar product of $|B^+_{ff}\rangle$ and $|k\rangle$ from finite size
    scaling (numerics) and comparison with CFT prediction. $h_k$
    is the conformal dimension of the field $|k\rangle$. Chains
    of size up to $L=30$ sites are used.}
 \label{tab:scal_prod_fix_scal}
\end{table}

\section{Loop models}
\label{sec:lattice_states}

\subsection{BCFT of loop models}
\label{sec:bcft_loops}

The lattice model we consider now is the dense loop model based on
the Temperley-Lieb algebra underlying the Potts model
\cite{Martin1991}.  We consider a system of $L$ (even or odd) strands,
with free (reflecting) boundary conditions at both boundaries.  At the critical
point, the anisotropic version of the model is defined by the
Hamiltonian:
\begin{equation}
  \label{eq:H_TL}
  H = -\sum_{i=0}^{L-2}e_i\, ,
\end{equation}
where $e_i$ are the TL generators. This Hamiltonian acts on link
states, patterns of connectivities of sites, and has a triangular block
form, with $j=0,2,\dots,L$ ($L$ even) or $j=1,3,\dots,L$ ($L$ odd),
the number of through lines, indexing each block \cite{Martin1991}.

The continuum limit of this loop model, when we parametrize the loop
weight as $\beta=2\cos(\pi/(p+1))$ ($p \geq 1$ is a real parameter
which we keep generic here), is a CFT with central charge
$c=1-6/(p(p+1))$.  The primary boundary fields of the resulting
(quasi-rational) boundary CFT are $\phi_j$, $j\in \mathbb{N}$ (see
e.~g.~\cite{Read2007}).  $\phi_j$ is interpreted as
the operator creating $j$ lines
in the continuum limit of the loop model, and corresponds to the
irreducible Virasoro representation with highest weight
$\phi_{1,1+j}$ (Kac table notation), with conformal dimension
$h_{1,1+j}=\frac{j(jp-2)}{4(p+1)}$.
The primaries $\phi_j$ generically fuse
as spin-($j/2$) su$(2)$ representations:
\begin{equation}
  \label{eq:fusion_phi}
  V_i \times V_j = \sum_{k\in \mathcal{I}} V_k\, , \quad 
  \mathcal{I}=\{|i-j|,|i-j|+2,\dots,i+j\}\, .
\end{equation}
The allowed boundary conditions one can put on the strip are given by
the labels $j$ of Virasoro representations.  Setting $i$ and $j$ on the
boundaries will then allow the propagation of only the sectors given
by the fusion of $i$ and $j$ \cite{Cardy1989}.  It is possible to 
manufacture microscopic boundary conditions for  the lattice model to 
reproduce this behavior \cite{Pearce2006}
(and see below).  Note that free (reflecting)
boundary conditions on the lattice,
without restrictions on the number of through lines, correspond to
setting on both boundaries the label $j\to\infty$.  The BCC operator
between boundaries $j$ and $j'$ is given by $\phi^{j,j'}_{|j-j'|}$,
the operator of smallest conformal weight in the spectrum of the
transfer matrix for a strip with $j$ and $j'$ on the sides. 
Fusion rules of these boundary operators are only a subset of those
for free boundaries, eq.~\eqref{eq:fusion_phi} (because
some boundary OPE structure constants vanish):
\begin{equation}
  \label{eq:phi_restr}
  \phi^{i,j}_{|i-j|} \times \phi^{j,k}_{|j-k|} = \sum_{s \in \mathcal{I}'}
  \phi_s^{i,k}\, .
\end{equation}
where $\mathcal{I}' = $ 
{\scriptsize $\left(
  \{|i-k|,|i-k|+2,\dots,i+k\}\cap
  \{||i-j|-|j-k||,||i-j|-|j-k||+2,\dots,|i-j|+|j-k|\}\right)$}. This
restriction comes from the compatibility of the number of lines
inserted by the fields in the corners and that which is allowed by the
boundary conditions.  We depict in figure \ref{fig:two_point_loops}
the geometric interpretation of the two-point function of BCC
operators as the contraction of $|i-j|$ lines originating from the
point in which the boundary condition is changed from $i$ to $j$ and
terminating at the point in which we change from $j$ to $i$.
More details about this geometric interpretation of the fields will be given
in the next section, where we also present numerical checks.

\begin{figure}[hpt]
\centering
\begin{tikzpicture}[]
  \draw (-1.75,0) node{$\langle \phi_{|i-j|}^{i,j}(x_1)
    \phi_{|i-j|}^{j,i}(x_2)\rangle$};
  \draw[<->] (0.5,0)--(1.5,0) ;
  \draw (2,0) -- (3,0);
  \draw[red] (3,0) -- (4,0);
  \draw (4,0) -- (5,0);
  \draw (3,0) to[out=90,in=180]
  (3.5,1) to[out=0,in=90] (4,0); 
  \draw (3,0) to[out=110,in=180]
  (3.5,1.2) to[out=0,in=60] (4,0); 
  \draw (3,0) to[out=60,in=180]
  (3.5,0.8) to[out=0,in=110] (4,0); 
  \draw[fill=black] (3,0) circle (0.04); 
  \draw[fill=black] (4,0) circle (0.04); 
  \draw (2.15,0.2) node{$i$};
  \draw (4.85,0.2) node{$i$};
  \draw (3.5,0.2) node{$j$};
  \draw (3,-0.2) node{$x_1$};
  \draw (4,-0.2) node{$x_2$};
  \draw[gray, thin] (3.5,0.65) -- (3.5,1.35);
  \draw (3.5,1.6) node{$|i-j|$};
\end{tikzpicture}
\caption{Geometric interpretation of the two-point function of BCC
  operators.}
    \label{fig:two_point_loops}
\end{figure}

\subsection{Numerics}
\label{sec:num}

We now want to understand which states on the lattice are described by
our basis boundary states introduced in section \ref{sec:sol_gluing}.
Define the normalized link state $\bsket{i}{j}{s}_L$ as:
\begin{equation}
  \label{eq:L_def}
  \bsket{i}{j}{s}_L:=
  \frac{1}{\sqrt{{}_L\langle{}^{i \phantom{a} j}_{\phantom{a} s} \bsket{i}{j}{s}_L}}
  \begin{tikzpicture}
    \begin{scope}[scale=0.5,dotdia/.style={cross out, draw,
        solid, red, inner sep=2pt}]
      \useasboundingbox (-1,-1) rectangle (12,1.5);
      \draw (-0.5,0) -- (11,0);
      \foreach \x in {0,1.5,9,10.5} 
      { 
        \draw (\x,0)--(\x,-2); 
      } 
      \foreach \x in {4,6} 
      { 
        \draw (\x,0) to[out=-90,in=180]
        (\x+0.25,-0.5) to[out=0,in=-90] (\x+0.5,0); 
      }
      \draw (3.5,0) to[out=-90,in=180]
      (3.5+1.75,-1) to[out=0,in=-90] (3.5+3.5,0); 
      \draw (2,0) to[out=-90,in=180]
      (2+3.25,-2) to[out=0,in=-90] (2+6.5,0); 
      \node at (0.75,-0.25) {$\cdots$};
      \node at (2+0.8,-0.25) {$\cdots$};
      \node at (5.3,-0.25) {$\cdots$};    
      \node at (9.8,-0.25) {$\cdots$};    
      \node at (7.8,-0.25) {$\cdots$};    
      \node at (2.75,0.75) {$n_c$};
      \node at (7.75,0.75) {$n_c$};
      \draw[thin,|-|] (2,0.5)--(3.5,0.5);
      \draw[thin,|-|] (7,0.5)--(8.5,0.5);
      \node at (0.75,0.75) {$n_l$};
      \node at (9.75,0.75) {$n_r$};
      \draw[thin,|-|] (0,0.5)--(1.5,0.5);
      \draw[thin,|-|] (9,0.5)--(10.5,0.5);
    \end{scope}
  \end{tikzpicture}
\end{equation}
with 
\begin{equation}
  \label{eq:sl}
  n_l+n_c=i\, ; \quad   
  n_r+n_c=j\, ; \quad   
  n_l+n_r=s\, .
\end{equation}
Here, $\sqrt{{}_L\langle{}^{i \phantom{a} j}_{\phantom{a} s} \bsket{i}{j}{s}_L}$
 is the lattice norm of the state.  It is obtained by using 
 the loop scalar product of link patterns \cite{Dubail2010}, which turns out to converge nicely to the 
 Virasoro bilinear form in the continuum
\cite{Dubail2010}.  
We call the $s=n_l+n_r$ lines not paired, through lines. The link
state we consider is projected onto the sector with exactly
$s$ through lines, 
and each time the through lines are contracted we get
zero. 
The $n_c$ lines inserted at
 the left corner are paired with the $n_c$ ones inserted at the right
corner, and during the evolution, these lines can propagate through
the system or be contracted.
Now we claim that this state flows in the continuum limit to
the following basis state:
\begin{equation}
  \label{eq:basis_states_lattice}
  \bsket{i}{j}{s}_L
  \overset{\mbox{\scriptsize{(FP)}}}{\longrightarrow} 
  \alpha_{ij}^s \bsket{i}{j}{s}\, .
\end{equation}
The limit to be taken is $L\to \infty$ with $i,j,s$ fixed, and we keep
only the finite part (noted as (FP)) which is supposed to be described
by the CFT, as will be discussed more precisely below. This
identification corroborates the interpretation of the boundary
operators $\phi_{j}$ and $\phi_{i}$ as inserting
respectively $j=n_r+n_c$ lines at the right corner and $i=n_l+n_c$ lines
at the left corner. We remark again that
the paired $n_c$ lines on the left corner can be
contracted with the $n_c$ lines on the right corner during the
evolution, but the remaining $s=n_r+n_l$ are forced to propagate in the
system.  Our discretization of the basis states is a natural
generalization of that for the case of no line insertion discussed in
\cite{Bondesan2011}.

We support our claim with the numerical results (labeled num) of
table \ref{tab:scal_prod_lattice}, obtained for fugacity of loops
$\beta=\sqrt{2}$ and to be compared with the CFT results (labeled
theo), corresponding to central charge $c=1/2$.  Denote by
$|k\rangle_L$ the $k$-th eigenvector (normalized according to the loop
scalar product) of the Hamiltonian restricted to the sector with
$s$ through lines. As discussed above, the
scalar product of a rectangular boundary state and $|k\rangle_L$ has
the form \eqref{eq:mlogBk}, and now $\gamma =
-\log\left(\tilde{\alpha}\right)$, $\tilde{\alpha}:=\alpha
4^{h_s-h_i-h_j}$, see eq.~\eqref{eq:ish_state}.  In the data presented
below $\tilde{\alpha}$ is determined by giving the expected value of
$a_1$ ($a_1$ theo) as input in the fit.  

The scaling limit $|k\rangle$ of $|k\rangle_L$ used to determine
$\langle B|k\rangle $ (theo), are for generic $k$ given by a
combination of Virasoro descendants at a level determined by the
energy of $|k\rangle$. The state $|0\rangle$ is clearly identified
with the primary $|\phi_j\rangle$ for each sector.  Then recall that
the continuum theory has field content given generically by the
quotient of Virasoro Verma modules $V_{1,1+s}/V_{1,-1-s}$ (with the
usual Kac table notation for Verma modules) \cite{Read2007}.
In the sector $s=0$, we identify then (due to normalization)
$|1\rangle=\sqrt{2/c}L_{-2}|0\rangle$, $|2\rangle = \sqrt{1/2c}
L_{-3}|0\rangle$. When $s=1$, we have $|1\rangle=1/\sqrt{2h_{1}}
L_{-1}|\phi_1\rangle$, $|2\rangle=1/\sqrt{4h_1+c/2}
L_{-2}|\phi_1\rangle$. For $s=2$, $|1\rangle=1/\sqrt{2h_2}
L_{-1}|\phi_2\rangle$, and $|2\rangle= 2/3 L_{-2}|\phi_2\rangle$ at
$c=1/2$ (note that this last is a priori an unknown combination of
Virasoro descendants at level two, which for $c=1/2$ is fixed using
that $|\phi_2\rangle$ is also degenerate at level two, since
$h_{1,3}=h_{2,1}$).  Scalar products $\langle B|k\rangle $ (theo) in
table \ref{tab:scal_prod_lattice} are then computed using the above
formulas for $|k\rangle$'s and the explicit form of $|B\rangle$'s,
eq.~\eqref{eq:ish_state} (see
Appendix \ref{sec:bs_two_one_leg} for $\bsket{1}{1}{0}$,
$\bsket{1}{1}{2}$ and eq.\eqref{eq:Baab_first_lev} for $\bsket{0}{1}{1}$ and
$\bsket{2}{0}{2}$), specialized at $p=3$.

\begin{table}[h!c]\tiny 
  \centering
  \begin{tabular}
    {cccccccc}
    \toprule
    $|B\rangle_L$ & $a_1$ num & $a_1$ theo & $\tilde{\alpha}_{ij}^s$ 
    & $\langle B|1\rangle$ num &   $\langle B|1\rangle$ theo &  
    $\langle B|2\rangle$ num &  $\langle B|2\rangle$ theo \\
    \midrule
    $\bsket{1}{1}{0}_L$
    & $0.18712(307)$ & $0.1875$ &
     $1.27053(16)$   
    & $1.50004(21)$  
    & $1.5$ & $0$
    & $0$\\
    \midrule
    $\bsket{0}{1}{1}_L$
    & $0.06278(131)$ & $0.0625$ 
    & $1.08322(6)$
    & $0.70704(5)$ & $\simeq 0.70711$
    & $0.35360(3)$ & $\simeq 0.35355$ \\
    \midrule
    $\bsket{2}{0}{2}_L$
    & $0.93778(865)$ & $0.9375$
    & $4.33940(152)$
    & $2.00057(74)$ & $-2$ 
    & $2.50616(105)$ & $2.5$ \\
    \midrule
    $\bsket{1}{1}{2}_L$   
    & $0.18938(76)$ & $0.1875$ & $1.79691(8)$ & 
    $0$ & $0$ & $0.50115(15)$ & $0.5$ \\
    \bottomrule
  \end{tabular}
  \caption{Lattice scalar products for loop fugacity $\beta=\sqrt{2}$
    and comparison with the results for the $c=1/2$ CFT.
    Extrapolations are determined by fitting values for
    $L=10\to 24$.
     $\tilde{\alpha}_{ij}^s=\alpha_{ij}^s 4^{h_s-h_i-h_j}$.
  }
 \label{tab:scal_prod_lattice}
\end{table}


Finally, we note that the  state defined in (\ref{eq:L_def}) is of course not the only microscopic state giving rise to $\bsket{i}{j}{s}$ in the continuum limit. Any similar state obtained via local modifications (e.~g.~by allowing a finite number of arches between the through lines, etc.) would obey the same property, albeit with a different value of the coefficient  $\alpha_{ij}^s$.

\subsection{Geometric interpretation of conformal blocks}
\label{sec:geom_conf_blocks}

\begin{figure}[hpt]
  \centering
  \begin{tikzpicture}[scale=0.5]
    \node at (-4,1) {$s=2:$};
    \draw[] (0,0) rectangle (3,2);
    \draw[thick] (0,2) -- (0,2.75);
    \draw[thick] (3,2) -- (3,2.75);
    \draw[thick] (0.5,2) to[out=90,in=-180]
    (0.5+1,2.5) to[out=0,in=90] (0.5+2,2); 
    \draw[thick] (0,0) -- (0,-0.75);
    \draw[thick] (3,0) -- (3,-0.75);
    \draw[thick] (0.5,0) to[out=-90,in=180]
    (0.5+1,0-0.5) to[out=0,in=-90] (0.5+2,0); 
    \draw[thick] (0,0) to[out=30,in=-95] (0.35,1) to[out=95,in=-30] (0,2); 
    \draw[thick] (3,0) to[out=90+30,in=-90+5] (3-0.35,1) to[out=85,in=-90-30] (3,2); 
    \draw[thick] (0+0.5,0) to[out=30,in=-95] (0.35+0.5,1) to[out=95,in=-30] (0+0.5,2); 
    \draw[thick] (3-0.5,0) to[out=90+30,in=-90+5] (3-0.35-0.5,1) to[out=85,in=-90-30] (3-0.5,2); 
    \draw[dashed] (-0.25,2.25)--(0.75,2.25);
    \draw[dashed] (-0.25,-0.25)--(0.75,-0.25);
    \node at (-1,2.25) {{\scriptsize $i=2$}};
    \node at (-1,-0.25) {{\scriptsize $j=2$}};
    \draw[dashed] (3.25,2.25)--(2.25,2.25);
    \draw[dashed] (3.25,-0.25)--(2.25,-0.25);
    \node at (4,2.25) {{\scriptsize $l=2$}};
    \node at (4,-0.25) {{\scriptsize $k=2$}};
    \node at (5.5,1) {$,$};
    \begin{scope}[xshift=8.5cm]
    \draw[] (0,0) rectangle (3,2);
    \draw[thick] (0,2) -- (0,2.75);
    \draw[thick] (3,2) -- (3,2.75);
    \draw[thick] (0.5,2) to[out=90,in=-180]
    (0.5+1,2.5) to[out=0,in=90] (0.5+2,2); 
    \draw[thick] (0,0) -- (0,-0.75);
    \draw[thick] (3,0) -- (3,-0.75);
    \draw[thick] (0.5,0) to[out=-90,in=180]
    (0.5+1,0-0.5) to[out=0,in=-90] (0.5+2,0); 
    \draw[thick] (0,0) to[out=30,in=-95] (0.35,1) to[out=95,in=-30] (0,2); 
    \draw[thick] (3,0) to[out=90+30,in=-90+5] (3-0.35,1) to[out=85,in=-90-30] (3,2); 
    \draw[thick] (0+0.5,0) to[out=80,in=-180-5] (1+0.5,0.75) to[out=0+5,in=90+10] (2+0.5,0); 
    \draw[thick] (3-0.5,2) to[out=270-10,in=0+5] (3-1-0.5,2-0.75) to[out=180-5,in=-80] (0.5,2); 
    \draw[dashed] (-0.25,2.25)--(0.75,2.25);
    \draw[dashed] (-0.25,-0.25)--(0.75,-0.25);
    \node at (-1,2.25) {{\scriptsize $i=2$}};
    \node at (-1,-0.25) {{\scriptsize $j=2$}};
    \draw[dashed] (3.25,2.25)--(2.25,2.25);
    \draw[dashed] (3.25,-0.25)--(2.25,-0.25);
    \node at (4,2.25) {{\scriptsize $l=2$}};
    \node at (4,-0.25) {{\scriptsize $k=2$}};
  \end{scope}
    \begin{scope}[xshift=5cm,yshift=-5cm]
    \node at (-4,1) {$s=4:$};
    \draw[] (0,0) rectangle (3,2);
    \draw[thick] (0,2) -- (0,2.75);
    \draw[thick] (3,2) -- (3,2.75);
    \draw[thick] (0.5,2) -- (0.5,2.75);
    \draw[thick] (3-0.5,2) -- (3-0.5,2.75);
    \draw[thick] (0,0) -- (0,-0.75);
    \draw[thick] (3,0) -- (3,-0.75);
    \draw[thick] (0.5,0) -- (0.5,-0.75);
    \draw[thick] (3-0.5,0) -- (3-0.5,-0.75);
    \draw[thick] (0,0) to[out=30,in=-95] (0.35,1) to[out=95,in=-30] (0,2); 
    \draw[thick] (3,0) to[out=90+30,in=-90+5] (3-0.35,1) to[out=85,in=-90-30] (3,2); 
    \draw[thick] (0+0.5,0) to[out=30,in=-95] (0.35+0.5,1) to[out=95,in=-30] (0+0.5,2); 
    \draw[thick] (3-0.5,0) to[out=90+30,in=-90+5] (3-0.35-0.5,1) to[out=85,in=-90-30] (3-0.5,2); 
    \draw[dashed] (-0.25,2.25)--(0.75,2.25);
    \draw[dashed] (-0.25,-0.25)--(0.75,-0.25);
    \node at (-1,2.25) {{\scriptsize $i=2$}};
    \node at (-1,-0.25) {{\scriptsize $j=2$}};
    \draw[dashed] (3.25,2.25)--(2.25,2.25);
    \draw[dashed] (3.25,-0.25)--(2.25,-0.25);
    \node at (4,2.25) {{\scriptsize $l=2$}};
    \node at (4,-0.25) {{\scriptsize $k=2$}};
  \end{scope}
  \end{tikzpicture}
\caption{Configurations contributing to the lattice
amplitudes ${}_L\langle {}^{2\,\,\, 2}_{\,\,\, s}| T^{L'} |
{}_{2\,\,\, 2}^{\,\,\, s}\rangle_L$ 
for $s=2,4$.}
\label{fig:conf_lines}
\end{figure}
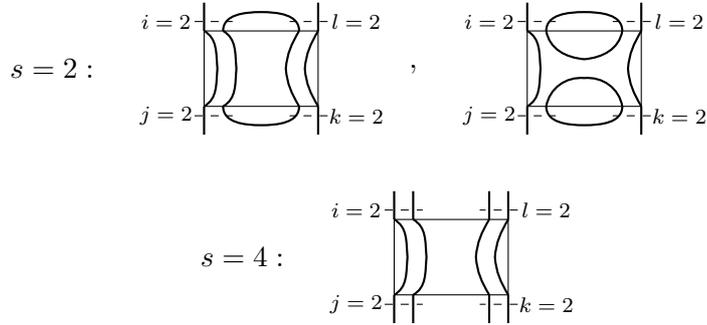

Up to now we have given numerical support to our claim that the
rectangle basis states $\bsket{i}{j}{s}$ defined in
eq.~\eqref{eq:ish_state} can be interpreted as the continuum limit of
microscopic boundary states with a given pattern of
connectivities. Here we want to push this discussion further and
associate a geometric picture in terms of loop configurations to
conformal blocks building the correlator of BCC operators.
According to the previous discussion, we have the identification:
\begin{equation}
  {}_L\bsbra{i}{l}{s} T^{L'} \bsket{j}{k}{s}_L 
  \overset{\mbox{\scriptsize{(FP)}}}{\longrightarrow} 
  \alpha_{jk}^s\alpha_{il}^s \mathcal{A}_{s}^{i,j,k,l}(\tau)\, .
\end{equation}
We depict in figure \ref{fig:conf_lines} two configurations contributing to the
lattice amplitude ${}_L\bsbra{2}{2}{2} T^{L'} \bsket{2}{2}{2}_L $ and
one to the amplitude ${}_L\bsbra{2}{2}{4} T^{L'} \bsket{2}{2}{4}_L $,
drawing only lines inserted at the corners.

We see that $\mathcal{A}_{s}^{i,j,k,l}(\tau)$ is then associated to
the continuum limit of lattice amplitudes where we insert $i$ lines in
top left corner, $l$ in top right, $j$ in the bottom left and $k$ in
bottom right.  $s$ of the lines (the most exterior ones) are forced to
propagate.  The other lines can---but are not obliged to---be contracted. Only if
$s=i+l=j+k$ ($s=4$ in the example of figure \ref{fig:conf_lines}), the
amplitude is associated to configurations where all lines inserted at
the corners are forced to propagate.  For example, the CFT partition
function of figure \ref{fig:part_fun_lines} will be
$\Zrect{i}{i}{i+j}{0}\propto L^w \mathcal{A}_{i+j}^{i,i,j,j}$. The
only term in the amplitude sensible to the changing of the number of
through lines (associated to different fusion sectors) is the
conformal block. Then we are led to the geometrical identification of
conformal blocks which we schematically represent as
\begin{center}
  \begin{tikzpicture}[scale=0.5]
    \node at (-3.25,1){$\mathcal{F}^s_{il;jk}(1-\zeta)
      \longleftrightarrow$};
    \draw (0,0) rectangle (3,2);
    \draw[thick] (0,0) to[out=0,in=-90]
    (0.5,1) to[out=90,in=0] (0,2); 
    \draw[thick] (3,0) to[out=180,in=-90]
    (2.5,1) to[out=90,in=180] (3,2); 
    \draw[thick] (3,0)--(2,0.5);
    \draw[thick] (3,2)--(2,1.5);
    \draw[thick] (0,0)--(1,0.5);
    \draw[thick] (0,2)--(1,1.5);
    \draw[fill=gray!40] (1.5,1) circle (0.65cm);
  \end{tikzpicture}
\end{center}
where the diagram on the right has a fixed number $s$ of through lines
and the bubble stays for the possibility of contracting or not the
other lines inserted at the corner. 

We recall that in general the
partition function $Z$ is given by a sum of amplitudes
from the relation \eqref{eq:Z_ampli2}, and 
its geometric interpretation follows from that of the amplitude.
We will comment more on that in section \ref{sec:gen_case},
where we discuss the universal character of the coefficients
$\alpha_{ij}^s$, and the precise relations between lattice
amplitudes and CFT partition functions \eqref{eq:Z_ampli}.

\subsubsection{Probabilistic interpretation}
\label{sec:prob_interp}

\begin{figure}[hpt]
\centering
\begin{tikzpicture}[]
  \draw (-1.75,0) node{$\mathcal{G}^2(1-\zeta)$};
  \draw[<->] (-0.75,0)--(-0.25,0) ;
\draw (0,-0.5) rectangle (1.5,0.5);
      \draw[thick] (0,-0.5) to[out=0,in=-90]
      (0.5,0) to[out=90,in=0] (0,0.5); 
      \draw[thick] (1.5,-0.5) to[out=180,in=-90]
      (1,0) to[out=90,in=180] (1.5,0.5); 
    \begin{scope}[xshift=6cm]
      \draw (-1.75,0) node{$\mathcal{G}^2(\zeta)$};
      \draw[<->] (-0.75,0)--(-0.25,0) ;
  \draw (0,-0.5) rectangle (1.5,0.5); 
  \draw[thick] (0,-0.5) to[out=90,in=180]
    (0.75,-0.1) to[out=0,in=90] (1.5,-0.5); 
  \draw[thick] (0,0.5) to[out=-90,in=180]
    (0.75,0.1) to[out=0,in=-90] (1.5,0.5); 
  \end{scope}
  \begin{scope}[yshift=-2cm]
  \draw (-1.75,0) node{$\mathcal{G}^0(1-\zeta)$};
  \draw[<->] (-0.75,0)--(-0.25,0) ;
      \draw (0,-0.5) rectangle (1.5,0.5);
      \draw[thick] (0,-0.5) to[out=0,in=-90]
      (0.5,0) to[out=90,in=0] (0,0.5); 
      \draw[thick] (1.5,-0.5) to[out=180,in=-90]
      (1,0) to[out=90,in=180] (1.5,0.5); 
    \begin{scope}[xshift=3cm]
      \draw (-0.75,0) node{$+\frac{1}{\beta}$};
      \draw (0,-0.5) rectangle (1.5,0.5); 
      \draw[thick] (0,-0.5) to[out=90,in=180]
      (0.75,-0.1) to[out=0,in=90] (1.5,-0.5); 
      \draw[thick] (0,0.5) to[out=-90,in=180]
      (0.75,0.1) to[out=0,in=-90] (1.5,0.5); 
  \end{scope}
  \end{scope}
\end{tikzpicture}
\caption{Geometric interpretation of the conformal block
  $\mathcal{G}^2$ and $\mathcal{G}^0$.  Thick lines represent the
  non-contractible lines created by the one-leg operators in the
  corners propagating or not through the system (time is flowing
  upwards).}
\label{fig:F1_F1_loops}
\end{figure}
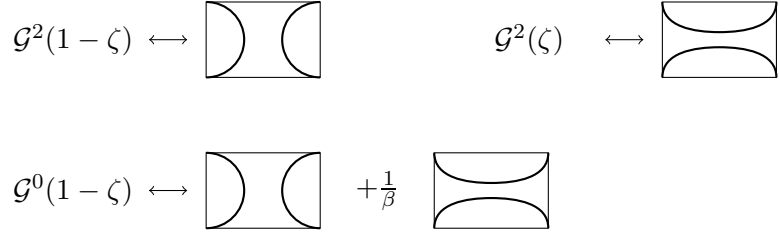

We can easily obtain
interesting results from the discussion above as follows.
We specialize to the
case of one line insertion at every corner, $i=j=k=l=1$.
In this case, the amplitudes are formally equivalent to that
for the Potts model with insertions of $\phi_{1,2}^{f,F}$ 
changing from free $f$ boundary conditions to fixed $F$, and
were computed explicitly 
in section \ref{sec:four_op}.
The partition function
\begin{equation*}
  \Zrect{1}{1}{2}{0}(\tau)=
  \mathcal{N}^2
  \eta^{-c/2}(\tau) \theta_3^{16 h}(\tau) \mathcal{G}^2(1-\zeta) 
  \propto L^w \mathcal{A}_2(\tau) 
\end{equation*}
is unambiguously the (universal part of the) continuum limit of
 a loop model where the two lines inserted at left and right
bottom corners
constrained to propagate in
the vertical direction.  Consequently
\begin{equation*}
  \Zrect{2}{0}{1}{1}(\tau)
=\mathcal{N}^2 L^{w} 
\eta^{-c/2}(\tau) \theta_3^{16 h}(\tau)
\mathcal{G}^2(\zeta(\tau))  
\propto (L')^{w} \mathcal{A}_2(-1/\tau)
\, ,
\end{equation*}
is then the partition function of a loop model with the two
lines inserted at left bottom and left top corners
constrained to propagate in the horizontal direction.  Then one
can interpret the conformal blocks
$\mathcal{G}^{a}$ as in the top of figure \ref{fig:F1_F1_loops}.  The
geometric interpretation of every other amplitude involving boundary
states with $\phi_{1}$ insertions at the corner will follow from the
decomposition in terms of $\mathcal{A}_2(\tau)$ and
$\mathcal{A}_2(-1/\tau)$.  For example for $\mathcal{A}_0(\tau)$,
we can use the hypergeometric identity $\mathcal{G}^0(1-\zeta)=1/F_{20}
\mathcal{G}^2(\zeta) - F_{22}/F_{20} \mathcal{G}^2(1-\zeta) \propto
\mathcal{G}^2(\zeta) +1/\beta \mathcal{G}^2(1-\zeta)$, see the bottom of
figure \ref{fig:F1_F1_loops}.

We have seen that the two elementary events which can happen when
inserting a line in each corner are described by 
$\mathcal{G}^2(1-\zeta)$ and $\mathcal{G}^2(\zeta)$. 
Then, the probability of having these two lines flowing
vertically is:
\begin{equation}
  \label{eq:prob_2_line_vert}
  \begin{tikzpicture}[]
  \useasboundingbox (0,-0.75) rectangle (10.75,0.5);
    \draw (-0.8,-0.265) node{$P(\tau) =$};
    \begin{scope}[xshift=1cm]
      \draw (0,0) rectangle (1,0.5);
      \draw[thick] (0,0) to[out=0,in=-90]
      (0.35,0.25) to[out=90,in=0] (0,0.5); 
      \draw[thick] (1,0) to[out=180,in=-90]
      (0.65,0.25) to[out=90,in=180] (1,0.5); 
    \end{scope}
    \begin{scope}[xshift=0cm,yshift=-1cm]
      \draw (0,0.75)--(3,0.75);
      \draw (0,0) rectangle (1,0.5);
      \draw[thick] (0,0) to[out=0,in=-90]
      (0.35,0.25) to[out=90,in=0] (0,0.5); 
      \draw[thick] (1,0) to[out=180,in=-90]
      (0.65,0.25) to[out=90,in=180] (1,0.5);
      \draw (1.5,0.25) node{$+$};
    \begin{scope}[xshift=2cm] 
      \draw (0,0) rectangle (1,0.5); 
      \draw[thick] (0,0) to[out=90,in=180]
      (0.5,0.2) to[out=0,in=90] (1,0); 
      \draw[thick] (0,0.5) to[out=-90,in=180]
      (0.5,0.3) to[out=0,in=-90] (1,0.5); 
    \end{scope}
    \end{scope}
    \draw (7.4,-0.265) node{$ 
      \displaystyle{=
        \frac{\Zrect{1}{1}{2}{0}(\tau)}{
          \Zrect{1}{1}{2}{0}(\tau)+\Zrect{2}{0}{1}{1}(\tau)}
        =
        \frac{\mathcal{G}^1(1-\zeta)}
        {\mathcal{G}^1(1-\zeta)+\mathcal{G}^1(\zeta)}}$};
  \end{tikzpicture}
{}
\end{equation}
The probability that the lines propagate horizontally is clearly
$1-P(\tau)$. Note that whatever normalization of the partition function
drops out since it is common to both partition functions.  

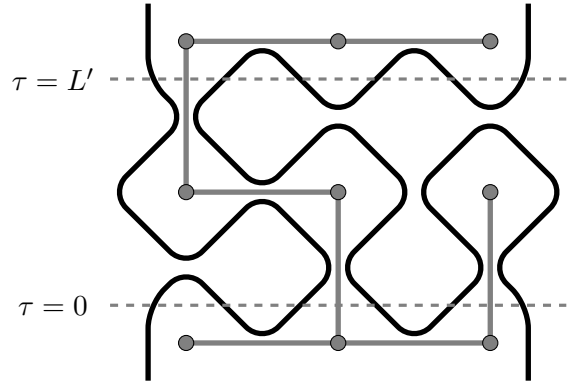
\begin{figure}[htp]
  \centering
  \begin{tikzpicture}[scale=1]
  \path (0,0) coordinate (n1);
  \path (2,0) coordinate  (n2);
  \path (4,0) coordinate (n3);
  \path (0,2) coordinate (n4);
  \path (2,2) coordinate  (n5);
  \path (4,2) coordinate (n6);
  \path (0,4) coordinate (n7);
  \path (2,4) coordinate  (n8);
  \path (4,4) coordinate (n9);
  \draw[line width=0.7mm,gray] (n1)--(n2)--(n3)--(n6);
  \draw[line width=0.7mm,gray] (n2)--(n5)--(n4)--(n7)--(n8)--(n9);
  \draw[line width=0.7mm, rounded corners=0.3cm]
  (-0.5,-0.5)-- ++(0,1)--
  ++(0.5,0.5)-- ++(1,-1)-- ++(1,1)-- ++(-1,1)-- ++(-1,-1)-- ++(-1,1)--   ++(1,1)-- ++(-0.5,0.5)-- ++(0,1);
  \draw[line width=0.7mm, rounded corners=0.3cm]
  (4.5,-0.5)-- ++(0,1)-- ++(-0.5,0.5)-- ++(1,1)-- ++(-1,1)-- ++(-1,-1)
  -- ++(1,-1)-- ++(-1,-1)--   ++(-1,1)-- ++(1,1)-- ++(-1,1)-- ++(-1,-1)-- ++(-1,1)-- ++(1,1)--  ++(1,-1)-- ++(1,1)-- ++(1,-1)-- ++(0.5,0.5)-- ++(0,1);
    \draw[very thick, dashed, black!50] (-1,0.5)--(5,0.5);
  \draw[very thick, dashed, black!50] (-1,3.5)--(5,3.5);
  \node at (-1.75,0.5) {$\tau=0$};
  \node at (-1.75,3.5) {$\tau=L'$};
  \draw[fill=gray] (n1) circle (0.1cm);
  \draw[fill=gray] (n2) circle (0.1cm);
  \draw[fill=gray] (n3) circle (0.1cm);
  \draw[fill=gray] (n4) circle (0.1cm);
  \draw[fill=gray] (n5) circle (0.1cm);
  \draw[fill=gray] (n6) circle (0.1cm);
  \draw[fill=gray] (n7) circle (0.1cm);
  \draw[fill=gray] (n8) circle (0.1cm);
  \draw[fill=gray] (n9) circle (0.1cm);
\end{tikzpicture}
  \caption{
    A configuration of loops (black) and the corresponding configuration of
    Fortuin-Kasteleyn clusters (gray).
    The through lines inserted at the bottom corners are forced
    to propagate to the top, and wired boundary conditions
    (all bonds open)
    are imposed on the links of the bottom 
    and top rows of the lattice where the Fortuin-Kasteleyn clusters
    live. $\tau=0$ and $\tau=L'$ are initial and final discrete
    imaginary times of the transfer matrix evolution in the vertical
    direction.}
  \label{fig:Sw_conf}
\end{figure}

We first consider the case of percolation, for which 
$\beta=1$, corresponding to $c=h=0$. We have:
\begin{align}
  \label{eq:Cardy_form}
  P_{c=0,\mbox{\scriptsize{dense}}}(\zeta)
  &=
  \frac{(1-\zeta)^{\frac{1}{3}}{}_2F_1\left(\frac{1}{3},\frac{2}{3};\frac{4}{3};1-\zeta\right)}
  {(1-\zeta)^{\frac{1}{3}}{}_2F_1\left(\frac{1}{3},\frac{2}{3};\frac{4}{3};1-\zeta\right) +
    \zeta^{\frac{1}{3}}{}_2F_1\left(\frac{1}{3},\frac{2}{3};\frac{4}{3};\zeta\right)}\\
  &=
  \frac{\Gamma(\smfrac{2}{3})}
    {\Gamma(\smfrac{4}{3})\Gamma(\smfrac{1}{3})}(1-\zeta)^{\frac{1}{3}}
  {}_2F_1\left(\frac{1}{3},\frac{2}{3};\frac{4}{3};1-\zeta\right)\\
  &=
  \frac{ \Gamma\left( \smfrac{2}{3}\right)}{ \Gamma\left(\smfrac{1}{3}\right) 
    \Gamma\left(\smfrac{4}{3}\right)}
  (1 - \zeta)^{\frac{1}{3}}
  \left(1
    - \frac{1}{6} (1 - \zeta)
    + \frac{5}{63} (1 - \zeta)^{2} 
    +O\left( (1-\zeta)^{3}\right)
  \right)
  \, .
\end{align}
This quantity is precisely the crossing probability computed by Cardy in
\cite{Cardy1992,Cardy2001}, that is, the probability that there is at
least one Fortuin-Kasteleyn (FK) crossing cluster spanning the rectangle in
the vertical direction. We now comment on this relation.
Microscopically, the configurations of loops contributing to the
numerator of our formula are those where the two lines inserted at the
corners propagate from bottom to top.  In terms of FK
clusters (which are encircled by the loops), this corresponds to
counting cluster configurations with ``wired'' boundary conditions on top
and bottom rows, and to the constraint that there is at least one
spanning cluster crossing the rectangle, see figure \ref{fig:Sw_conf}.
Call such ensemble of cluster configurations $\mathcal{S}_w$.

Instead the configurations counted by
Cardy's formula, which we denote by  $\mathcal{S}$,
 are simply those with
at least one spanning cluster, without imposing wired boundary conditions.
We now show that 
\begin{equation}
  \label{eq:S_Sw}
  \#(\mathcal{S})=2^{L}\times 2^L\times 
  \#(\mathcal{S}_w)\, ,
\end{equation}
where $\#(E)$ is the number of elements of the set $E$.
First note that changing the status (open or closed) of links in
the bottom and top rows of the lattice where FK clusters live,
 transforms a configuration
$s\in \mathcal{S}$ into another $s' \in \mathcal{S}$. Then we can
group the spanning configurations in classes differing by only the
status of bottom and top links, and for each class we choose the
representative with all the bottom and top links open. Since there are
$2^L\times 2^L$ (first factor for the possible status of top links and
second for those of bottom links) elements in each class,
eq.~\eqref{eq:S_Sw} follows.  
Now note that also the set of all possible (not only spanning)
cluster configurations can be divided in
classes with elements differing by the status of top and bottom links as before.
Then, if we compute the crossing probability by
dividing the number of spanning
configurations $\#(\mathcal{S})$
by the number of all possible configurations, 
the multiplicative factor $2^L\times 2^L$ drops out,
and the equality of our formula with Cardy's one is established.  The
above argument does not work anymore when we have
generic weight of clusters $\beta$, since spanning
configurations cannot be regrouped then due to the different powers of
$\beta$ weighting each configuration.  Crossing probabilities for 
generic fugacity $\beta$ have been computed using SLE methods, see
\cite{Bauer2006}.


\subsection{Logarithmic cases}

We have anticipated in section \ref{sec:four_op} that for certain
(logarithmic) values of the central charge the conformal blocks 
for $\phi_1$ insertions at the corners,
\eqref{eq:conf_block_0_phi12}-\eqref{eq:conf_block_1_phi12}, are
ill-defined. Indeed, more fundamentally fusion of the fields at these
points cannot be decomposed onto irreducible Virasoro modules
\cite{Gaberdiel2001,Gurarie2004}.  For the operators degenerate at
level two we are considering, this happens for $c=-2$ and
$\phi_1=\phi_{1,2}$ $h=-1/8$ (dense polymers), and for $c=0$ and
$\phi_1=\phi_{2,1}$, $h=5/8$ (dilute polymers). In these cases, we
have the following behavior of conformal blocks ($\epsilon=p-1$
in \eqref{eq:conf_blocks_c-2}, $\epsilon=p-2$ in
\eqref{eq:conf_blocks_c0}, where we parametrized $c=1-6/(p(p+1))$
\begin{align}
  \label{eq:conf_blocks_c-2}
  \lim_{\epsilon\to 0}
  \mathcal{G}^{0}(\zeta) &= \frac{2}{\pi} K(\zeta)\, ; \qquad
  \lim_{\epsilon\to 0}
  \mathcal{G}^{2}(\zeta) = \frac{2}{\pi} K(\zeta)\, ; \\
  \label{eq:conf_blocks_c0}
  \mathcal{G}^{0}(\zeta) &= 
  \frac{15}{16 \epsilon} S_2(\zeta) +O(\epsilon)  \, ; \qquad
  \lim_{\epsilon\to 0}
  \mathcal{G}^{2}(\zeta) = S_2(\zeta)\, . \qquad
\end{align}
$K$ is the complete elliptic integral of the first kind of parameter
$\zeta$ and 
\begin{equation}
  S_2(\zeta):=\zeta^2 \,
  _2F_1\left(-\frac{1}{2},\frac{3}{2};3;\zeta\right)\, .
\end{equation}
For dense
polymers the degeneracy 
$\mathcal{G}^{0}(\zeta)\to \mathcal{G}^{2}(\zeta)$
arises because the roots of the indicial
equation \eqref{eq:roots_ind} coincide, so that one has to replace one
conformal block with the missing solution of the differential equation
which generally is given by a derivation procedure w.r.t. the root,
and here can be written as ($\epsilon=p-1$)
\begin{equation}
  \lim_{\epsilon\to 0} \frac{\mathcal{G}^{2}(\zeta)-\mathcal{G}^{0}(\zeta)}{\epsilon}
  \propto K(1-\zeta) - \frac{\log(16)}{\pi} K(\zeta)\, .
\end{equation}
In terms of fusion rules this is stated as the degeneracy of the two
fields $\phi_0$ and $\phi_2$, which are replaced by the identity and
its logarithmic partner under the Virasoro algebra
\cite{Gaberdiel2001}.  For dilute polymers, $\mathcal{G}^{0}$ here
has to be replaced by the logarithmic conformal block, which
can be written as ($\epsilon=p-2$)
\begin{align}
  \lim_{\epsilon\to 0} \frac{1}{F_{02}}
  \left(
    \mathcal{G}^{0}(\zeta)
    -
    F_{00}\mathcal{G}^{0}(1-\zeta)
  \right)
  &=
  (1-\zeta)^2 \, _2F_1\left(-\frac{1}{2},\frac{3}{2};3;1-\zeta\right)\\
  &=S_2(1-\zeta)\, ,
\end{align}
where $F_{ij}$ are the fusing matrices of four $\phi_{1}$ fields at
generic central charge, eq.~\eqref{eq:Fij}.
Again this is stated as the degeneracy of the stress tensor
and the field $\phi_2$, which are organized in an indecomposable
module under the Virasoro algebra \cite{Gurarie2004}.

The presence of a divergence is clearly a difficulty for the
interpretation of the partition function as that of a statistical
model. We consider the partition functions $\Zrect{1}{1}{a}{0}$,
$a=0,2$ computed in section \ref{sec:four_op}, for which we have the geometrical
interpretation discussed in the previous section. In particular recall
that $\Zrect{1}{1}{2}{0}$ is given by configurations of loops
where we constrain the two lines inserted at the bottom corners
to propagate without being contracted.
We assume now that the geometric interpretation of boundary states
goes through also in the dilute case, where the value of 
of $h$ in the formulas is $h_{2,1}$. 
We expect physically that these partition functions
should be well defined for every value of loop fugacity
$\beta$, also at $\beta=0$, which is the case we are after
for polymers.  The solution to this is simply hidden in the
normalization of the partition function, eq.~\eqref{eq:Zfinal_N}.
In the case of dense polymers $\mathcal{N}^0=\mathcal{N}^2=0$.
This does not come as a surprise, and should be traced back to the presence
of a single loop on the rectangle.
Factorizing this trivial zero, one then gets that for dense polymers:
\begin{align}
  \label{eq:Z02111_c-2}
  \Zrect{1}{1}{2}{0}(h=-1/8)=\Zrect{1}{1}{0}{0}(h=-1/8)&=
  \sqrt{L}
  \eta \theta_3^{-2}\frac{2}{\pi} K(1-\zeta)\\
  &=
  \sqrt{L} \eta\, ,
\end{align}
where we used the identity $2K(1-\zeta(\tau))=\pi \theta_3(\tau)^2$.
As a highly not-trivial check of the above result we can compare with
the expression coming from the determinant of the Laplacian with
different boundary conditions on opposite edges, see eq.~\eqref{eq:NNDD}.

For dilute polymers
one has 
\begin{align}
  \mathcal{N}^0 &\to \frac{\pi}{2}\epsilon + O(\epsilon^2)\\
  \mathcal{N}^1 &\to \frac{15}{32}\pi + O(\epsilon)\, .
\end{align}
Note that the indecomposability parameter $b=5/6$ \cite{Gurarie2004},
characterizing the Jordan cell of the stress tensor
for dilute polymers,  is
present explicitly in the coefficients, $h^2/b=15/32$.  Finally we
predict again degeneracy of both partition functions
\begin{align}
  \label{eq:Z02111_c0}
  \Zrect{1}{1}{2}{0}(h=5/8)=\Zrect{1}{1}{0}{0}(h=5/8)&=
  L^{-5}
  \theta_3^{-10} S_2(1-\zeta)\, .  
\end{align}
The degeneracy we have observed could be understood more fundamentally
as the gluing of the two sectors $\phi_0$ and $\phi_2$ in an
indecomposable module corresponding to a single boundary condition.


It is of interest now to see what is the behavior of probabilities
\eqref{eq:prob_2_line_vert} at the logarithmic points discussed above,
$c=-2$ dense and $c=0$ dilute. The outcome is that probabilities are
well defined but their expansion shows logarithmic terms. 
For expanding around $\zeta=1$, the case of a tall rectangle,
 we use
\begin{equation}
  \mathcal{G}^2(\zeta)=F_{20}\mathcal{G}^0(1-\zeta)+F_{22}\mathcal{G}^2(1-\zeta) 
  \, .
\end{equation}
As $h\to-1/8$ (dense case) we see from formulas \eqref{eq:Fij} that
$F_{20}\to\infty$ and $F_{22}\to-\infty$, but
$F_{20}+F_{22}\to\log(16)/\pi$, and recall that the conformal
blocks are degenerate, eq.~\eqref{eq:conf_blocks_c-2}.
The probability reads:
\begin{align}
  \label{eq:P_den_poly}
  P_{c=-2,\mbox{\scriptsize{dense}}}(\zeta) 
  &=
  \frac{K(1-\zeta)}{K(1-\zeta)+K(\zeta)} \\
  &=
  \frac{\pi }{\pi-\log \left(\frac{1-\zeta}{16}\right)}
  +
  \frac{\pi(1-\zeta)}
  {2\left(\pi-\log \left(\frac{1-\zeta}{16}\right) \right)^2} 
  +
  O\left((1-\zeta)^2\right)\, .
\end{align}
When instead $h\to 5/8$ (dilute case) we have $F_{20}\to b/h^2=32/15$ and
$F_{22}\to-\infty$.   In this case the divergence of
$\mathcal{G}^0$, eq.~\eqref{eq:conf_blocks_c0},
compensates that of the OPE structure constant, and the result
is finite:
\begin{align}
  \label{eq:P_dil_poly}
  P_{c=0,\mbox{\scriptsize{dilute}}}(\zeta) 
  &=
  \frac{S_2(1-\zeta)}{S_2(1-\zeta)+S_2(\zeta)}
  \\
  &=
  \frac{15 \pi }{32} (1-\zeta)^2
  +\frac{15 \pi}{32} (1-\zeta)^3\\
  &\quad+
  \frac{75 \pi \left(12 \log
    \left(\frac{1-\zeta}{16}\right)+12 \pi -1\right)}{4096} (1-\zeta)^4
  +
  O\left((1-\zeta)^5 \right)\, .
\end{align}

We are not aware of a previous derivation of the above probability
formulas.
Note that since $P(1-\zeta)=1-P(\zeta)$, the same series plus the
constant term holds for a very long rectangle ($\zeta\to 0$) 
replacing $1-\zeta$ by $\zeta$.  
We have already seen how rectangular amplitudes contain the OPE
structure constants, and now we see that for logarithmic CFTs the
indecomposability parameters (contained in the $F$-matrices)
 show up in geometric observables.

Note also that expressing the anharmonic ratio in terms of the modular
parameter $q$ through eq.~\eqref{eq:anratio}
\begin{equation}
  \label{eq:log_zeta_t}
  \log(1-\zeta) = \log (16)
  +\log (\sqrt{q})
  -8 \sqrt{q}+\dots
\end{equation}
integer powers of $\tau$ enter explicitly the expression of the
probability. This feature is a consequence of the Jordan form of the
transfer matrix in these cases.

More generally one could replace the ill-behaved conformal blocks in
the logarithmic cases by a combination of them which has probabilistic
interpretation and which is expected to be well defined even if the
conformal blocks themselves are singular or degenerate.  Unfortunately
writing down explicitly such a probabilistic basis in the space of
conformal blocks does not seem to be simple for higher number of line
insertions.  Indeed we do not know how to associate precisely a
conformal block to a given propagation of lines
(see discussion at the beginning of
 section \ref{sec:geom_conf_blocks}), but more importantly,
we do not know how to fix the normalization.


\subsection{OPE coefficients and lattice crossing symmetry}
\label{sec:gen_case}

Given a CFT rectangle boundary state $| B^b_{ac} \rangle= \sum_s
C_{ijs}^{abc}\sqrt{C_{ss0}^{aca}} \bsket{i}{j}{s}$, and the
discretization of basis states introduced above in
eq.~\eqref{eq:L_def}, we can define the more general lattice boundary
state as combination of lattice basis states
\begin{equation}
  \label{eq:bstate_lattice}
  | B^b_{ac} \rangle_L = \sum_s D_{ijs}^{abc} \bsket{i}{j}{s}_L\, .
\end{equation}
We expect then that $| B^b_{ac} \rangle_L  \overset{\mbox{\scriptsize{(FP)}}}{\longrightarrow}  | B^b_{ac} \rangle$ if
\begin{equation}
  \label{eq:D_C_alpha}
  D_{ijs}^{abc}:=C_{ijs}^{abc}\sqrt{C_{ss0}^{aca}} \frac{1}{\alpha_{ij}^s}\, .
\end{equation}
We now question the universal character of $\alpha$.  As the crossing
symmetry of correlation functions (or rectangular amplitudes) is a
powerful tool for constraining OPE structure constants in the
continuum theory, we will see that also lattice crossing symmetry
allows to derive constraints for the coefficients $D$'s appearing in
eq.~\eqref{eq:bstate_lattice} and to infer about the universality of the
coefficients $\alpha$'s.

Let us see concretely how to implement lattice crossing symmetry for
the simplest non-trivial case of $i=j=1$, corresponding to boundary
fields inserting one line in each corner.  For this situation all the
possible CFT rectangle boundary states one has to consider are:
\begin{align}
  \label{eq:Biip1ip12}
  |B_{i, i+2}^{i+1}\rangle &= 
  C_{1,1,2}^{i,i+1,i+2}\sqrt{C_{220}^{i,i+2,i}}
  \bsket{1}{1}{2}\, ,\\
  \label{eq:Biiipm12}
  |B_{i, i}^{i\pm 1}\rangle &= 
  C_{1,1,0}^{i,i\pm 1,i}\sqrt{C_{000}^{iii}}\bsket{1}{1}{0}
  +
  C_{1,1,2}^{i,i\pm 1,i}\sqrt{C_{220}^{iii}}\bsket{1}{1}{2}\, ,
\end{align}
where boundary labels $\in \mathbb{N}$.  The discretization of the
above basis boundary states follows from the discussion at the beginning
of section \ref{sec:num} :
\begin{align}
  \label{eq:101_disc} 
  \frac{1}{\beta^{L/4}} 
  \begin{tikzpicture}[scale=0.5]
    \useasboundingbox (-1,-0.2) rectangle (3.5,1);
      \draw (-0.75,0) -- (3.25,0);
      \foreach \x in {0,2} 
      { 
        \draw (\x,0) to[out=-90,in=180]
        (\x+0.25,-0.5) to[out=0,in=-90] (\x+0.5,0); 
      }
      \draw (-0.5,0) to[out=-90,in=180]
      (-0.5+1.75,-1) to[out=0,in=-90] (-0.5+3.5,0); 
      \node at (1+0.3,-0.25) {$\cdots$};
    \end{tikzpicture}
    &  \overset{\mbox{\scriptsize{(FP)}}}{\longrightarrow}  
    \alpha_{11}^0 \bsket{1}{1}{0} \, ,\\
  \frac{1}{\beta^{L/4-1/2}} 
  \begin{tikzpicture}[scale=0.5]
    \useasboundingbox (-1,-0.2) rectangle (3.5,1);
      \draw (-0.75,0) -- (3.25,0);
      \foreach \x in {0,2} 
      { 
        \draw (\x,0) to[out=-90,in=180]
        (\x+0.25,-0.5) to[out=0,in=-90] (\x+0.5,0); 
      }
      \draw (-0.5,0)--(-0.5,-1); 
      \draw (3,0)--(3,-1); 
      \node at (1+0.3,-0.25) {$\cdots$};
    \end{tikzpicture}
    &\overset{\mbox{\scriptsize{(FP)}}}{\longrightarrow} 
    \alpha_{11}^2 \bsket{1}{1}{2} \, .
\end{align}
Accordingly we introduce the discretizations
\begin{align}
  \label{eq:B_{i,i+1}^{i+1/2}L}
  |B_{i, i+2}^{i+1}\rangle_L &= 
  \frac{D_{112}^{i,i+1,i+2}}{\beta^{L/4-1/2}} 
  \begin{tikzpicture}[scale=0.5]
    \useasboundingbox (-1,-0.2) rectangle (3.25,1);
      \draw (-0.75,0) -- (3.25,0);
      \foreach \x in {0,2} 
      { 
        \draw (\x,0) to[out=-90,in=180]
        (\x+0.25,-0.5) to[out=0,in=-90] (\x+0.5,0); 
      }
      \draw (-0.5,0)--(-0.5,-1); 
      \draw (3,0)--(3,-1); 
      \node at (1+0.3,-0.25) {$\cdots$};
    \end{tikzpicture}
    \, ,\\
  \label{eq:B_{i,i}^{i+1/2}L}
  |B_{i, i}^{i\pm 1}\rangle_L &=
  \frac{  D_{110}^{i,i\pm 1,i}}{\beta^{L/4}}
  \begin{tikzpicture}[scale=0.5]
    \useasboundingbox (-1,-0.2) rectangle (11.5,1);
      \draw (-0.75,0) -- (3.25,0);
      \foreach \x in {0,2} 
      { 
        \draw (\x,0) to[out=-90,in=180]
        (\x+0.25,-0.5) to[out=0,in=-90] (\x+0.5,0); 
      }
      \draw (-0.5,0) to[out=-90,in=180]
      (-0.5+1.75,-1) to[out=0,in=-90] (-0.5+3.5,0); 
      \node at (1+0.3,-0.25) {$\cdots$}; \node at (5.25,0.05) {$+
        \displaystyle{\frac{D_{112}^{i,i\pm 1,i}}{\beta^{L/4-1/2}}}$};
      \begin{scope}[xshift=8.25cm]
        \draw (-0.75,0) -- (3.25,0);
        \foreach \x in {0,2} 
        { 
          \draw (\x,0) to[out=-90,in=180]
          (\x+0.25,-0.5) to[out=0,in=-90] (\x+0.5,0); 
        }
        \draw (-0.5,0)--(-0.5,-1); 
        \draw (3,0)--(3,-1); 
        \node at (1+0.3,-0.25) {$\cdots$};
      \end{scope}
    \end{tikzpicture}
    \, .
\end{align}

\vspace{0.2cm}

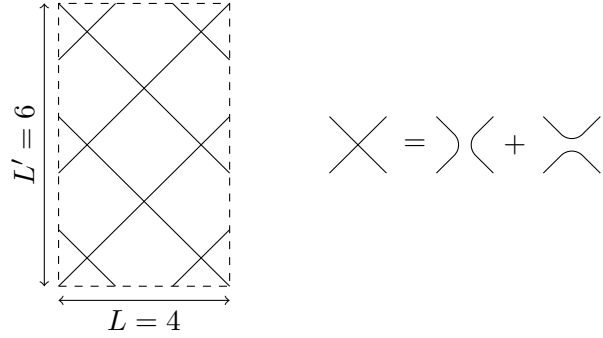
\begin{figure}[hpt]
\centering
\begin{tikzpicture}[scale=0.75]
  \draw[dashed] (0,0) rectangle (3,5);
  \foreach \y in {0,2}
  {
    \draw (0,\y)--(3,\y+3);
    \draw (3,\y)--(0,\y+3);
  }
  \draw (0,4)--(1,5);
  \draw (2,5)--(3,4);
  \draw (0,1)--(1,0);
  \draw (2,0)--(3,1);

  \draw[<->] (0,-0.25)--(3,-0.25) node[pos=0.5,below] {$L=4$};
  \draw[<->] (-0.25,0)--(-0.25,5) node[pos=0.5,above,rotate=90] {$L'=6$};
  \begin{scope}[xshift=5cm,yshift=2cm,scale=0.5]
    \begin{scope}[xshift=-0.5cm]
      \draw[postaction={decorate}] (2,2)--(1,1);
      \draw[postaction={decorate}] (1,1)--(2,0);
      \draw[postaction={decorate}] (0,0)--(1,1);
      \draw[postaction={decorate}] (1,1)--(0,2);
    \end{scope}             
    \node at (2.5,1) {$=$}; \draw (3.25,0) [rounded corners=0.2cm] --
    (4.25,1) -- (3.25,2); \draw (5.25,2) [rounded corners=0.2cm] --
    (4.25,1) -- (5.25,0); \node at (6,1) {$+$}; \draw (7,0) [rounded
    corners=0.2cm] -- (8,1) -- (9,0); \draw (9,2) [rounded
    corners=0.2cm] -- (8,1) -- (7,2); 
  \end{scope}
  
\end{tikzpicture}
\caption{A lattice of size $(L=4)\times (L'=6)$. At each node
we have the transfer matrix depicted on the right, equal to
identity plus TL generators.}
\label{fig:LLplattice}
\end{figure}

Now consider an $L\times L'$ lattice as in figure
\ref{fig:LLplattice}, where each vertex corresponds to a node transfer
matrix equal to the identity plus the Temperley Lieb generator.

\begin{figure}[hpt]
\centering
\begin{tikzpicture}[scale=0.75]
  \draw[dashed] (0,0) rectangle (3,5);
  \foreach \y in {0,2}
  {
    \draw (0,\y)--(3,\y+3);
    \draw (3,\y)--(0,\y+3);
  }
  \draw (0,4)--(1,5);
  \draw (2,5)--(3,4);
  \draw (0,1)--(1,0);
  \draw (2,0)--(3,1);
  \foreach \y in {1,3}
  {
    \draw (0,\y) to[out=180,in=-90] (0-0.5,\y+0.5) to[out=90,in=180] (0,\y+1);
    \draw (3,\y) to[out=0,in=-90] (3+0.5,\y+0.5) to[out=90,in=0] (3,\y+1);
  }
  \draw[fill=gray!30] (0,0) rectangle (3,-1);
  \draw (1.5,-0.5) node{$|B_{i,i}^{i+1}\rangle_L$};
  \draw[fill=gray!30] (0,5) rectangle (3,6);
  \draw (1.5,5.5) node{$|B_{i,i}^{i\pm1}\rangle_L$};
  \draw (4,2.5) node{$=$};
  \begin{scope}[xshift=6cm]
  \draw[dashed] (0,0) rectangle (3,5);
  \foreach \y in {0,2}
  {
    \draw (0,\y)--(3,\y+3);
    \draw (3,\y)--(0,\y+3);
  }
  \draw (0,4)--(1,5);
  \draw (2,5)--(3,4);
  \draw (0,1)--(1,0);
  \draw (2,0)--(3,1);
  \foreach \x in {1}
  {
    \draw (\x,0) to[out=-90,in=180] (\x+0.5,0-0.5) to[out=0,in=-90] (\x+1,0);
    \draw (\x,5) to[out=90,in=180] (\x+0.5,5+0.5) to[out=0,in=90] (\x+1,5);
  }
  \draw[fill=gray!30] (0,0) rectangle (-1,5);
  \draw[fill=gray!30] (3,0) rectangle (4,5);
  \draw (-0,2.4999)--(-0,2.5) node[pos=0.5,above,sloped] {$|B_{i\pm
      1,i+1}^{i}\rangle_{L'}$};
  \draw (3,2.5)--(3,2.4999) node[pos=0.5,above,sloped] {$|B_{i\pm
      1,i+1}^{i}\rangle_{L'}$};

  \end{scope}
\end{tikzpicture}
\caption{Equation imposing lattice crossing symmetry.}
\label{fig:modinv_lattice}
\end{figure}
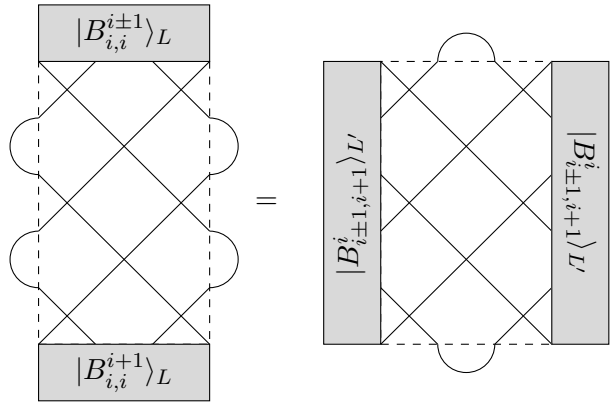

A partition function on this lattice is specified by giving an initial
and a final state, and computing the loop scalar product of the final
state with the sum over all possible choices of node operators acting
on the initial state.  The universal part of the partition function in
the continuum limit is a partition function in boundary CFT.  The
conformal boundary conditions are imposed by the choice of boundary
states according to the identification of their continuum limit
as discussed in section  \ref{sec:num}.
Clearly one can decide to consider time flowing from bottom to top or
from left to right; the equality of these two descriptions for each
configuration is a constraint that we call lattice crossing symmetry
and which allows for the determination of (ratios of) the
coefficients $D$'s. See figure \ref{fig:modinv_lattice} for a picture
of the resulting equation.

Consider first the case of boundary conditions $i,i+1,i,i-1$ around
the rectangle. Taking the identity operator acting in the bottom-top
direction (or TL generators in the left-right direction) at every
vertex, gives the equation
\begin{equation}
  \label{eq:id_bot_top_cis2}
  {}_L\langle B_{i,i}^{i-1}|B_{i,i}^{i+1}\rangle_L
  = 
  \beta^{L/2-1} {}_{L'}\langle B_{i+1,i-1}^{i}|\prod_{i=0}^{L'/2-1}
  e_{2i+1}|B_{i+1,i-1}^{i}\rangle_{L'} = 0\, ,
\end{equation}
leading to the first relation:
\begin{equation}
  \label{eq:ciii+1/2_ciii-1/2}
  D_{110}^{i,i+1,i}  D_{110}^{i,i-1,i}=-
  D_{112}^{i,i+1,i}  D_{112}^{i,i-1,i}\, .
\end{equation}
Then consider the case with $i,i+1,i,i+1$ around the rectangle.
Again taking identity in the bottom-top direction at every vertex
gives
\begin{equation}
  \label{eq:id_bot_top_cis3}
  {}_L\langle B_{i,i}^{i+1}|B_{i,i}^{i+1}\rangle_L
  = 
  \beta^{L/2-1} {}_{L'}\langle B_{i+1,i+1}^{i}|\prod_{i=0}^{L'/2-1}
  e_{2i+1}|B_{i+1,i+1}^{i}\rangle_{L'} \, ,
\end{equation}
yielding:
\begin{equation}
  \label{eq:ciii+1/2_A}
  \left(D_{110}^{i,i+1,i}\right)^2+
  \left(D_{112}^{i,i+1,i}\right)^2
  =
  \beta^{L/2-L'/2}\beta \left(D_{110}^{i+1,i,i+1}\right)^2\, .
\end{equation}
Now, taking instead contractions in the bottom-top direction (identity
for left-right direction) at every vertex gives
\begin{equation}
  \label{eq:id_bot_top_cis} 
  \beta^{L'/2-1} {}_{L}\langle B_{i,i}^{i+1}|\prod_{i=0}^{L/2-1}
  e_{2i+1}|B_{i,i}^{i+1}\rangle_{L}
 = {}_{L'}\langle B_{i+1,i+1}^{i}|B_{i+1,i+1}^{i}\rangle_{L'}
 \, ,
\end{equation}
yielding:
\begin{equation}
  \label{eq:D1_D2_rel2}
  \left(D_{110}^{i+1,i,i+1}\right)^2+
  \left(D_{112}^{i+1,i,i+1}\right)^2
  =
  \beta^{L'/2-L/2}\beta \left(D_{110}^{i,i+1,i}\right)^2\, .
\end{equation}
Defining 
\begin{equation}
  \label{eq:C_D1_D2}
  \mathcal{C}_{\pm}^{i}=\frac{D_{112}^{i,i\pm 1,i}}{D_{110}^{i,\pm 1,i}}\, ,
\end{equation}
these equations imply:
\begin{equation}
  \label{eq:ciii+1/2_ci+1/2i+1/2i}
  \mathcal{C}_{+}^{i} = 
  \sqrt{\frac{\beta^2}{1 + \left(\mathcal{C}_{-}^{i+1}\right)^2}
    -1}\, , \quad \mathcal{C}_{+}^{i}=-\frac{1}{\mathcal{C}_{-}^{i}}\, .
\end{equation}
These are recursion relations for the coefficients $\mathcal{C}$ with
initial condition $\mathcal{C}_{+}^{0}=0$, and are independent of the
system sizes $L,L'$.  The solution of the recursion can be expressed
in terms of Chebyshev polynomials of the second kind $U_n(x)$ as:
\begin{align}
  \label{eq:Cpm_U}
  \mathcal{C}_{\pm}^i = \pm
  \sqrt{\frac{U_{i\mp 1}\left(\frac{\beta}{2}\right)}
    {U_{i\pm 1}\left(\frac{\beta}{2}\right)}}\, .
\end{align}
The first few coefficients read explicitly
\begin{align}
  \mathcal{C}_{-}^1&= \sqrt{\beta^2-1}\, ; \quad
  \mathcal{C}_{+}^1= -\frac{1}{\sqrt{\beta^2-1}}\, ,\\
  \mathcal{C}_{-}^2 &= \sqrt{\beta^2-2}\, ; \quad
  \mathcal{C}_{+}^2 = -\frac{1}{\sqrt{\beta^2-2}}\, ,\\
  \mathcal{C}_{-}^3 &= \sqrt{\frac{\beta^4-3 \beta^2 +
      1}{\beta^2-1}}\, ; \quad \mathcal{C}_{+}^3 =
  -\sqrt{\frac{\beta^2-1}{\beta^4-3 \beta^2 + 1}}\, , \dots
\end{align}
We now claim that these expressions are not restricted to the two
cases solved above and indeed solve every possible constraint one can
write down from switching the left-right and bottom-top descriptions
of lattice partition functions. 
Using \eqref{eq:D_C_alpha} and setting $i=1$, we have that the ratio
$\alpha_{11}^2/\alpha_{11}^0$ is:
\begin{align}
  \label{eq:alpha_ratio}
  \frac{\alpha_{11}^2}{\alpha_{11}^0} = 
  \frac{1}{\sqrt{\beta^2-1}}
  \frac{C_{112}^{101}\sqrt{C_{220}^{111}}}{C_{110}^{101}\sqrt{C_{000}^{111}}}\, .
\end{align}
Note that crossing symmetry only allows to determine $\alpha$'s up to
a common constant. 

As already remarked, if we had started with lattice boundary states
differing from those we have used by local modifications 
(e.g.~by allowing a finite number of arches between the through lines, etc.), 
we would expect that the continuum limit would be the same, 
but with different $\alpha$'s. For example, we could consider
instead of eq.~\eqref{eq:101_disc}:
\begin{align}
  \frac{1}{\beta^{L/4}} 
  \begin{tikzpicture}[scale=0.5]
    \useasboundingbox (-1,-0.2) rectangle (4.5,1);
      \draw (-0.75,0) -- (4.25,0);
      \foreach \x in {-0.5,1,3} 
      { 
        \draw (\x,0) to[out=-90,in=180]
        (\x+0.25,-0.5) to[out=0,in=-90] (\x+0.5,0); 
      }
      \draw (0.5,0) to[out=-90,in=180]
      (0.5+1.75,-1) to[out=0,in=-90] (0.5+3.5,0); 
      \node at (2+0.3,-0.25) {$\cdots$};
    \end{tikzpicture}
      \overset{\mbox{\scriptsize{(FP)}}}{\longrightarrow}  
    \hat{\alpha}_{11}^0 \bsket{1}{1}{0} \, .
\end{align}
If we start with different boundary states, the equations resulting
from lattice crossing symmetry would be modified in general (e.g.~if
we use the boundary state above, eq.~\eqref{eq:ciii+1/2_A} will have
an extra factor of $\beta^2$ in the r.~h.~s.~, similarly
eq.~\eqref{eq:D1_D2_rel2}).  The ratio of
coefficients for boundary states with local modifications
$\hat{\alpha}_{11}^2/\hat{\alpha}_{11}^0$ should always be independent
of system size and related to $\beta$ and OPE constants, but in
general different from \eqref{eq:alpha_ratio}.  Since we expect more
generally that $\alpha_{ij}^s$ 
changes under a local modification of the microscopic state
(e.g.~by extra factors of $\beta$ in the example above), it
cannot be predicted by the CFT.  Note that the ``physical'' states
$|B_{ac}^b\rangle_L$ defined in eq.~\eqref{eq:bstate_lattice} are
normalized in order to obtain in the continuum the CFT states
$|B_{ac}^b\rangle$, and the CFT partition functions \eqref{eq:Z_ampli}, when
taking overlaps.


The expression \eqref{eq:alpha_ratio} can be simplified further using
eq.~\eqref{eq:cons_rect_F} for $i=j=k=l=1$, $a=c=0$, $b=d=1$, $r=2$:
\begin{equation}
  \label{eq:C_F_10}
  (C_{110}^{101})^2 C_{000}^{111}  F_{02} {\scriptsize \left[ \begin{array}{cc}
        1 & 1 \\
        1 & 1 \end{array} \right]}
  +
  (C_{112}^{101})^2 C_{220}^{111}F_{22}{\scriptsize \left[ \begin{array}{cc}
        1 & 1 \\
        1 & 1 \end{array} \right]}
  = 0\, ,
\end{equation}
and the explicit results \eqref{eq:Fij}
yield
\begin{align}
  \label{eq:a1/a0}
  \frac{\alpha_{11}^2}{\alpha_{11}^0} = 
  \sqrt{
    \frac{\beta 
      \Gamma \left(-\smfrac{8 h_1}{3}-\smfrac{1}{3}\right) \Gamma \left(\smfrac{2}{3}-\smfrac{8
   h_1}{3}\right)   }
    {(\beta^2-1)
      \Gamma \left(\smfrac{1}{3}-\smfrac{4 h_1}{3}\right) \Gamma (-4 h_1)
    }}\, .
\end{align}

In table \ref{tab:alphas} we check these results against numerical
values obtained for $p=3,4,5$, with $p$
parametrizing the fugacity of loops as $\beta=2\cos(\pi/(p+1))$.
Large sizes which we did not study would be needed to get more precise
results, but nonetheless we find good agreement in these cases.

\begin{table}[h!c] 
  \centering
  \begin{tabular}
    {ccc}
    \toprule
    $p$ & num & theo \\
    \midrule
    $3$ & $0.70715(9)$ & $\simeq 0.70711$\\
    \midrule
    $4$ & $0.68506(32)$ & $\simeq 0.69267$\\
    \midrule
    $5$ & $0.66671(66)$ & $\simeq 0.68736$\\
    \bottomrule
  \end{tabular}
  \caption{Comparison of the ratio $\alpha_{11}^2/\alpha_{11}^0$   
    determined numerical (num) by fitting values for $L=10\to 24$ and
    theoretically (theo) using eq.~\eqref{eq:a1/a0} for different values of
    $p$, parametrizing the weight of loops 
    $\beta=2\cos(\pi/(p+1))$.}
 \label{tab:alphas}
\end{table}

Now that we have determined the ratio $\alpha_{11}^2/\alpha_{11}^0$
we can invert the reasoning and determine the OPE coefficients
in terms of it from eq.~\eqref{eq:Cpm_U}:
\begin{equation}
  \label{eq:C_a0_a1}
  \frac{C_{112}^{i,i\pm 1,i}\sqrt{C_{220}^{iii}}}
  {C_{110}^{i,i\pm 1,i}\sqrt{C_{000}^{iii}}}
  =
   \pm
   \sqrt{
    \frac{\beta 
      \Gamma \left(-\smfrac{8 h_1}{3}-\smfrac{1}{3}\right) \Gamma \left(\smfrac{2}{3}-\smfrac{8
   h_1}{3}\right)   }
    {(\beta^2-1)
      \Gamma \left(\smfrac{1}{3}-\smfrac{4 h_1}{3}\right) \Gamma (-4 h_1)
    }}
  \sqrt{\frac{U_{i\mp 1}\left(\frac{\beta}{2}\right)}
    {U_{i\pm 1}\left(\frac{\beta}{2}\right)}}\, .
\end{equation}
In particular the rectangle partition functions for $\phi_{1}$
insertion in each corner can be explicitly written as
(the functions $\mathcal{G}$'s are defined in section \ref{sec:four_op})
\begin{align}
  \label{eq:Zi,i+1/2,i+1,i+1/2}
  \begin{tikzpicture}[]
   \useasboundingbox (-0.3,0.2) rectangle (2,1.5);
    \draw (0,0) -- (1,0); 
    \draw (0,0.6) -- (1,0.6); 
    \draw (0,0) -- (0,0.6);
    \draw (1,0.6) -- (1,0); 
    \draw[fill=black] (0,0.6) circle (0.04); 
    \draw[fill=black] (1,0.6) circle (0.04); 
    \draw (-0.15,0.3) node{$i$};
    \draw (1.5,0.3) node{$i+2$};
    \draw[fill=black] (0,0) circle (0.04); 
    \draw[fill=black] (1,0) circle (0.04); 
    \draw (0.5,0.8) node{$i+1$};
    \draw (0.5,-0.2) node{$i+1$};
  \end{tikzpicture}
  &= A_i
  L^{c/4-8h_1}
  \eta^{-c/2} \theta_3^{16 h_1}
  \mathcal{F}^2(1-z)\\
  \begin{tikzpicture}[]
   \useasboundingbox (-0.3,0.2) rectangle (1.5,1.5);
    \draw (0,0) -- (1,0); 
    \draw (0,0.6) -- (1,0.6); 
    \draw (0,0) -- (0,0.6);
    \draw (1,0.6) -- (1,0); 
    \draw[fill=black] (0,0.6) circle (0.04); 
    \draw[fill=black] (1,0.6) circle (0.04); 
    \draw (-0.15,0.3) node{$i$};
    \draw (1.25,0.3) node{$i$};
    \draw[fill=black] (0,0) circle (0.04); 
    \draw[fill=black] (1,0) circle (0.04); 
    \draw (0.5,0.8) node{$i+1$};
    \draw (0.5,-0.2) node{$i+1$};
  \end{tikzpicture}
  &=B_i
  L^{c/4-8h_1}
  \eta^{-c/2} \theta_3^{16 h_1}
  \left(
  \mathcal{F}^0(1-z)
  +
  F_{02}
  \frac{\beta}{\beta^2-1}
  \frac{U_{i-1}\left(\frac{\beta}{2}\right)}
    {U_{i+1}\left(\frac{\beta}{2}\right)}
  \mathcal{F}^2(1-z)
  \right) \, .
\end{align}
The constant of proportionality $A_i$, $B_i$ can be read off from the
general formula \eqref{eq:Z_ampli2} and are given in terms of boundary
OPE coefficients. We note that for the CFT describing loop models we
are considering here, these coefficients are known and can be obtained
by analytic continuation of the $F$-matrices for minimal models
$\mathcal{M}(p,p+1)$ \cite{Furlan1990}.



\section{Conclusion}
\label{sec:conc}

In this work we have constructed CFT boundary states
encoding rectangular geometries with different boundary conditions, 
and elucidated their relation with correlators of boundary
conditions changing operators, completing the
investigation initiated in \cite{Bondesan2011}.  We have developed in
particular the geometrical interpretation in terms of lattice loop
models, and we have derived new formulas for
probabilities of self-avoinding walks.  Remarkably, these formulas
contain the indecomposability parameters of logarithmic CFTs in the
expansion of the amplitudes.

In conclusion, the  formalism developed in \cite{Bondesan2011} and the present paper gives access to large classes of loop models partition functions on the rectangle. A further generalization should allow one, for instance, to obtain closed formulas for the average conductance of rectangles at the higher plateau transitions in the spin quantum Hall effect, along the lines of \cite{Bondesan2011a,Bondesan2012}. 

\bigskip

\noindent {\bf Acknowledgments}: We thank J. Dubail for useful
discussions and earlier collaboration on a related work. This work was
supported in part by a grant from the ANR Projet 2010 Blanc SIMI 4:
DIME.

\appendix

\section{Useful formulas}
\label{sec:usef_form}
We collect in this appendix definitions and useful properties
of special functions appearing in the main part of the manuscript.

The theta functions we use are defined as:
\begin{align}
  \label{eq:eta}
  \eta(\tau) &= q^{1/24}\prod_{n=1}^\infty(1-q^n) =
  \left(\frac{\theta_2(\tau)\theta_3(\tau)\theta_4(\tau)}{2}\right)^{1/3}\\
  \label{eq:theta2}
  \theta_2(\tau) &= \sum_{n\in\mathbb{Z}}q^{(n+1/2)^2/2} 
  =2\frac{\eta(2\tau)^2}{\eta(\tau)}
  \\
  \label{eq:theta3}
  \theta_3(\tau) &= \sum_{n\in\mathbb{Z}}q^{n^2/2} 
  =\frac{\eta(\tau)^5}{\eta(\tau/2)^2\eta(2\tau)^2}
  \\
  \label{eq:theta4}
  \theta_4(\tau) &= \sum_{n\in\mathbb{Z}}(-)^n q^{n^2/2}
  =\frac{\eta(\tau/2)^2}{\eta(\tau)} \, ,
\end{align}
with $q=e^{2\pi i \tau}$, $\tau=L'/L$.  Denote by $K$ the complete elliptic
integral of the first kind with modulus $k$ and $K':=K(1-k^2)$ its
complementary, then $\tau = iK'/(2 K)$. One has
\begin{equation}
  \label{eq:K_theta}
  k = 
\left(\frac{\theta_2(2\tau)}{\theta_3(2\tau)}\right)^2 =
  \frac{\theta_3(\tau)^2-\theta_4(\tau)^2}{\theta_3(\tau)^2+\theta_4(\tau)^2}
  = \left(\frac{\pi\theta_2(\tau)^2}{4K}\right)^2
  ; \quad
  K = \frac{\pi}{4}(\theta_3(\tau)^2 + \theta_4(\tau)^2)\, .
\end{equation}

Modular properties of these theta functions are:

\begin{align}
  \eta(-1/\tau)&=\sqrt{-i\tau}\eta(\tau)\\
  \theta_2(-1/\tau)&=\sqrt{-i\tau}\theta_4(\tau)\\
  \theta_3(-1/\tau)&=\sqrt{-i\tau}\theta_3(\tau)\\
  \theta_4(-1/\tau)&=\sqrt{-i\tau}\theta_2(\tau)\, ,
\end{align}
\begin{align}
  \eta(\tau+1)&=e^{i\pi/12}\eta(\tau)\\
  \theta_2(\tau+1)&=e^{i\pi/4}\theta_2(\tau)\\
  \theta_3(\tau+1)&=\theta_4(\tau)\\
  \theta_4(\tau+1)&=\theta_3(\tau)\, .
\end{align}

\section{Boundary states for two one-leg insertions}
\label{sec:bs_two_one_leg}

We give here explicit expressions of basis 
boundary state $\bsket{1}{1}{0}$ and $\bsket{1}{1}{2}$
for generic $c=1-6/(p(p+1))$ up to level $8$:
\tiny
\begin{equation}
  \label{eq:B_lev8}
  \begin{split}
    &4^{2h_{1}}\bsket{1}{1}{0}
    =
    |0\rangle
    + 
    \left(7-\frac{24}{p+3}\right) L_{-2} |0\rangle  \\
    &+
    \left(-\frac{40}{3 p+5}+\frac{24}{p+3}-\frac{1}{2}\right)
    L_{-4} |0\rangle 
    +
    \left(\frac{200}{9 p+15}-\frac{48}{p+3}+\frac{19}{6}\right) 
    L_{-2}^2 |0\rangle \\     
    &-\frac{128 (p+1) p^2}{3 (p+3) (3 p+5) (5 p+7)}L_{-3}^2 |0\rangle 
    +\frac{(p (p (709
      p+769)-1665)+315)}{6 (p+3) (3 p+5) (5 p+7)} L_{-4}L_{-2} |0\rangle\\
    &+
    \frac{(p (667-p (103
      p+331))-105)}{6 (p+3) (3 p+5) (5 p+7)}
    L_{-2}^3 |0\rangle
    +
    \frac{64 (2 p-1) p}{3 (3 p+5) (5 p+7)} L_{-6} |0\rangle \\  
    &+ 
    \frac{(p-9) (p (p (709 p+481)-1761)+315)}{12 (p+3) (3 p+5) (5 p+7) (7 p+9)}
    L_{-8} |0\rangle 
    +
    \frac{64 p \left(p \left(6 p^2+p-16\right)+9\right)}{(p+3) (3 p+5) (5 p+7) (7 p+9)}
    L_{-6}L_{-2} |0\rangle \\
    &+
    \left(- 
      \frac{80}{3 p+5}-\frac{196}{5 p+7}+\frac{324}{7 p+9}+\frac{24}{p+3}+\frac{1}{8} 
    \right)L_{-4}L_{-2}^2 |0\rangle \\
    &+ 
    \left(\frac{800}{9 (3 p+5)}-\frac{1372}{3 (5 p+7)}+\frac{2916}{7 (7
      p+9)}+\frac{4}{p+3}-\frac{11}{504}\right) L_{-2}^4 |0\rangle \\
  &-\frac{256 p^2 (p+1)}{3 (3 p+5) (5 p+7) (7 p+9)} 
  L_{-5}L_{-3} |0\rangle \\
  &+
  \left(\frac{520}{9 (3 p+5)}+\frac{14896}{15 (5 p+7)}-\frac{8424}{7 (7
      p+9)}-\frac{60}{p+3}+\frac{53}{1260}\right) L_{-4}L_{-2}^2|0\rangle\\
    &-\frac{128 (p-9) p^2 (p+1)}{3 (p+3) (3 p+5) (5 p+7) (7 p+9)}
    L_{-3}^2L_{-2}|0\rangle
    +
    \dots \, ,
  \end{split}
\end{equation}
\tiny
\begin{equation}
  \label{eq:B_phi13}
  \begin{split}
    &4^{2h_{1}-h_2}
    \bsket{1}{1}{2}
    =
    |\phi_2\rangle  
    +
    \frac{5 p-1}{3 p+1} L_{-2} |\phi_2\rangle  
    -
    \frac{2 (p+1)}{3 p+1}L_{-1}^2 |\phi_2\rangle  \\
    &+
    \frac{34 p^3+21 p^2+44 p-3}{60 p^3+86 p^2+40 p+6} L_{-4}|\phi_2\rangle  
    -
    \frac{16 (p-2) p (p+1)}{(2 p+1) (3 p+1) (5 p+3)} 
    L_{-3}L_{-1}|\phi_2\rangle  \\
    &-
    \frac{p^2+34 p-3}{30 p^2+28 p+6} L_{-2}^2 |\phi_2\rangle  
    +
    \frac{20 p^3+58 p^2+44 p+6}{30 p^3+43 p^2+20 p+3} 
    L_{-2}L_{-1}^2 |\phi_2\rangle  
    -
    \frac{2 (p+1)^2 (2 p+3)}{30 p^3+43 p^2+20 p+3} 
    L_{-1}^4 |\phi_2\rangle \\
    &+   
    \frac{64 p (p (p (p (43 p+96)+44)-21)+18)}{9 (2 p+1) (3 p+1) (3 p+2) (5 p+3) (7 p+5)}
    L_{-6} |\phi_2\rangle  \\
    &+ 
    \frac{16 p (p+1) (p (p (17 p-9)-48)+252)}{9 (2 p+1) (3 p+1) (3 p+2) (5 p+3) (7 p+5)} L_{-5,}L_{ -1} |\phi_2\rangle  \\
    &+
\frac{p (p (p (p (8662 p+12621)+1175)-11631)-5337)+270}{18 (2 p+1) (3 p+1) (3 p+2) (5 p+3) (7
   p+5)}
 L_{-4}L_{ -2} |\phi_2\rangle \\ 
    &-\frac{128 p^2 (p (5 p+19)+6)}{9 (3 p+1) (3 p+2) (5 p+3) (7 p+5)}
    L_{-3}^2 |\phi_2\rangle  \\
    &+
    \frac{p (p (p (65-p (970 p+2159))+2997)+1833)+90}{3 (2 p+1) (3 p+1) (3 p+2) (5 p+3) (7 p+5)} L_{-4}L_{-1}^2 |\phi_2\rangle  \\
    &-
    \frac{16 p (p+1) (p (p (123 p-203)+144)+180)}{9 (2 p+1) (3 p+1) (3 p+2) (5 p+3) (7 p+5)} L_{-3}L_{-2}L_{-1} |\phi_2\rangle  
    +
    \frac{p (269-p (97 p+29))-15}{6 (3 p+1) (5 p+3) (7 p+5)}
    L_{-2}^3 |\phi_2\rangle  \\
    &+
    \frac{32 p (p+1)^2 (p (35 p+52)+60)}{9 (2 p+1) (3 p+1) (3 p+2) (5 p+3) (7 p+5)} L_{-3}L_{-1}^3 |\phi_2\rangle \\
    &+
    \frac{(p+1) (p (p (518 p-151)-921)-90)}{9 (2 p+1) (3 p+1) (3 p+2) (7 p+5)}
    L_{-2^2}L_{-1}^2 |\phi_2\rangle \\
    &-\frac{2 (p+1)^2 (p (7 p (58 p+1)-723)-270)}{9 (2 p+1) (3 p+1) (3 p+2) (5 p+3) (7 p+5)} L_{-2}L_{-1}^4 |\phi_2\rangle \\
    &+
    \frac{4 (p+1)^3 (p (10 p-17)-30)}{9 (2 p+1) (3 p+1) (3 p+2) (5 p+3) (7 p+5)}
    L_{-1}^6 |\phi_2\rangle 
    +
    \dots
\end{split}
\end{equation}
\normalsize

%
%
\bibliography{bs_loops}

\end{document}